\documentclass[reprint,amssymb,amsmath,aps,superscriptaddress,floatfix,pra,longbibliography]{revtex4-2}

\usepackage{graphicx}
\usepackage{amssymb}
\usepackage{epstopdf}
\usepackage{upgreek}
\usepackage{dcolumn}
\usepackage{bm}
\usepackage{gensymb}
\usepackage{braket}
\usepackage{amsmath}
\usepackage{verbatim}
\usepackage{mathdots}
\usepackage{float}
\usepackage{sidecap}
\usepackage{amsmath}
\usepackage[T1]{fontenc}

\usepackage{upgreek}

\expandafter\ifx\csname package@font\endcsname\relax\else
 \expandafter\expandafter
 \expandafter\usepackage
 \expandafter\expandafter
 \expandafter{\csname package@font\endcsname}%
\fi

\newcommand{\micron}{~$\upmu$m}

\newcommand{\be}{\begin{eqnarray}}
\newcommand{\ee}{\end{eqnarray}}
\newcommand{\bfig}{\begin{figure}}
\newcommand{\efig}{\end{figure}}

\begin{document}

\title{Precision Measurement of the Microwave Dielectric Loss of Sapphire \\ in the Quantum Regime with Parts-per-Billion Sensitivity}

\author{Alexander P. Read}
\thanks{These authors contributed equally to this work.\\ alex.read@yale.edu \\ benjamin.chapman@yale.edu}
\affiliation{Departments of Physics and Applied Physics, Yale University, New Haven, Connecticut 06511, USA}
\affiliation{Yale Quantum Institute, Yale University, New Haven, Connecticut 06511, USA}
\author{Benjamin J. Chapman}
\thanks{These authors contributed equally to this work.\\ alex.read@yale.edu \\ benjamin.chapman@yale.edu}
\affiliation{Departments of Physics and Applied Physics, Yale University, New Haven, Connecticut 06511, USA}
\affiliation{Yale Quantum Institute, Yale University, New Haven, Connecticut 06511, USA}
\author{Chan U Lei}
\affiliation{Departments of Physics and Applied Physics, Yale University, New Haven, Connecticut 06511, USA}
\affiliation{Yale Quantum Institute, Yale University, New Haven, Connecticut 06511, USA}
\author{Jacob C. Curtis}
\affiliation{Departments of Physics and Applied Physics, Yale University, New Haven, Connecticut 06511, USA}
\affiliation{Yale Quantum Institute, Yale University, New Haven, Connecticut 06511, USA}
\author{Suhas Ganjam}
\affiliation{Departments of Physics and Applied Physics, Yale University, New Haven, Connecticut 06511, USA}
\affiliation{Yale Quantum Institute, Yale University, New Haven, Connecticut 06511, USA}
\author{Lev Krayzman}
\affiliation{Departments of Physics and Applied Physics, Yale University, New Haven, Connecticut 06511, USA}
\affiliation{Yale Quantum Institute, Yale University, New Haven, Connecticut 06511, USA}
\author{Luigi Frunzio}
\affiliation{Departments of Physics and Applied Physics, Yale University, New Haven, Connecticut 06511, USA}
\affiliation{Yale Quantum Institute, Yale University, New Haven, Connecticut 06511, USA}
\author{Robert J. Schoelkopf}
\affiliation{Departments of Physics and Applied Physics, Yale University, New Haven, Connecticut 06511, USA}
\affiliation{Yale Quantum Institute, Yale University, New Haven, Connecticut 06511, USA}

\begin{abstract}
Dielectric loss is known to limit state-of-the-art superconducting qubit lifetimes. Recent experiments imply upper bounds on bulk dielectric loss tangents on the order of 100 parts-per-billion, but because these inferences are drawn from fully fabricated devices with many loss channels, 
these experiments do not definitely implicate or exonerate the dielectric.
To resolve this ambiguity, we devise a measurement method capable of separating and resolving bulk dielectric loss with a sensitivity at the level of $5\times10^{-9}$. The method, which we call the dielectric dipper, involves the \textit{in situ} insertion of a dielectric sample into a high-quality microwave cavity mode. Smoothly varying the sample participation in the cavity mode enables a differential measurement of the sample's dielectric loss tangent. The dielectric dipper can probe the low-power behavior of dielectrics at cryogenic temperatures and does so without the need for any lithographic process, enabling controlled comparisons of substrate materials and processing techniques.  
We demonstrate the method with measurements of sapphire grown by edge-defined film-fed growth (EFG) in comparison to high-grade sapphire grown by the heat-exchanger method (HEMEX). For EFG sapphire 
we infer a bulk loss tangent of $63(8)\times10^{-9}$ and a substrate-air interface loss tangent of $15(3)\times10^{-4}$~(assuming a sample surface thickness of $3~\text{nm}$). For a typical transmon, this bulk loss tangent would limit device quality factors to $Q\lessapprox 20 \times 10^6$, suggesting that bulk loss is likely the dominant loss mechanism in the longest-lived transmons on sapphire.  
We also demonstrate this method on HEMEX sapphire and bound its bulk loss tangent to be
less than $19(6)\times10^{-9}$. As this bound is about 3 times smaller than the bulk loss tangent of EFG sapphire, use of HEMEX sapphire as a substrate would lift the bulk dielectric coherence limit of a typical transmon qubit to several milliseconds. 
\end{abstract}
\maketitle

\section{Introduction}
\label{sec:Intro}

Superconducting circuits are a promising hardware platform for quantum information science, prized for their union of rapid gates and long coherence times.  That success is partly due to a two-decade effort which has prolonged the nanosecond-scale coherence of early superconducting qubits~\cite{nakamura:1999} by six-orders of magnitude~\cite{place:2021,wang:2022,gordon:2022,somoroff:2021}.

Some of these gains have come from engineering matrix-elements~\cite{manucharyan:2009} and insensitivity to decoherence mechanisms like 1/f noise~\cite{koch:2007}. Other improvements have been made by directly minimizing noise-spectral densities~\cite{paik:2011}, for example by improving the structure, materials, or fabrication process of the device. For either approach, the first step in finding the next order-of-magnitude improvement is to determine the dominant loss mechanism.

Identification of the dominant source of loss is complicated by the many loss channels present in superconducting qubits~\cite{mcrae:2020_materials}. These losses add together to limit the quality factor $Q$ of the qubit.
\begin{eqnarray}
\frac{1}{Q} = \frac{1}{Q_\text{conductor}} + \frac{1}{Q_\text{dielectric}} + \frac{1}{Q_\text{radiative}} + ...
\end{eqnarray}
While loss $Q^{-1}$ is straightforward to measure, it is unclear how to best improve qubit coherence without knowing which of these mechanisms is the dominant source of loss.
To overcome this challenge and distinguish between various sources of loss, one can measure $Q^{-1}$ on either a suite of devices, each designed to be more or less sensitive to particular loss channels, or a single device with some \textit{in situ} tunable experimental parameter, which preferentially aggravates or alleviates one source of dissipation.

Studies on dielectric loss have principally employed the suite approach~\cite{gao:2008,wang:2009,wang:2015,dial:2016,chu:2016,gambetta:2016,calusine:2018,woods:2019,mcrae:2020_dielectric}.  For transmons on sapphire with $Q \leq 4 \times 10^6$, those studies measured $Q^{-1}$ that scaled proportionally with the fraction of the qubit energy stored in surface dielectrics~\cite{wang:2015,dial:2016}.  This directly implicated surface dielectric loss as the dominant loss mechanism in these devices, motivating the use of fabrication processes compatible with more aggressive surface cleaning.  The result was some of the highest published transmon lifetimes to-date~\cite{place:2021,wang:2022}. 

Advances in coherence make it necessary to reassess which loss channels limit coherence. Quality factors of transmon qubits have now increased to $Q\approx 10^7$~\cite{place:2021,wang:2022,gordon:2022}. 
Does bulk dielectric loss limit the coherence of these devices?

A desirable technique to answer this question would uniquely distinguish bulk loss from other sources of loss in the system, and do so with high resolution. Specifically, it would have sensitivity to loss tangents much smaller than those which would limit transmon coherence, which in current devices would be a bulk loss tangent of $\text{tan}\delta_{\text{bulk}}\approx10^{-7}$.
It would also complement previous methodologies~\cite{calusine:2018,woods:2019} by measuring both power-dependent and power-independent losses without requiring lithography or other processing. Such flexibility would allow measurements of materials for which there is not yet an established process for depositing and patterning films and Josephson junctions. Similarly, standard substrates could be studied mid-process to determine the process stage at which losses manifest.

Here we introduce a suitable technique for this task, designed for the study of dielectric loss at microwave frequencies, cryogenic temperatures, and low powers, which allows dielectric loss to be measured independently from other qubit loss-channels. The technique, which we call the dielectric dipper, involves the \textit{in situ} insertion of a dielectric sample into a high-quality microwave cavity mode.  Smooth variation of the sample participation enables a differential measurement of the dielectric loss of the sample, with a sensitivity of $5 \times 10^{-9}$.

We use this technique to measure sapphire grown by edge-defined film-fed growth (EFG), and determine its bulk and substrate-air interface loss tangents to be $63(8)\times10^{-9}$, and $15(3)\times10^{-4}$, respectively \cite{tSA}.
We also measure high-grade sapphire grown by the heat-exchanger method (HEMEX) and bound the bulk loss tangent  to below $19(6)\times10^{-9}$.  These measurements suggest that current record-lifetime transmon qubits fabricated on EFG sapphire are approaching the bulk dielectric loss limit, and that the use of HEMEX sapphire as a substrate would suppress that loss, 
lifting the transmon coherence limit from bulk dielectric loss to several milliseconds. 

\section{Measurement technique}
\label{sec:Measurement technique}
\subsection{Overview}

One way to improve the resolution of a dielectric loss measurement is to heavily suppress the non-dielectric loss channels.
This consideration suggests the use of high-$Q$ microwave cavities as sensitive probes; the properties that make them promising as quantum memories---relatively few physical loss mechanisms and lifetimes from milliseconds to seconds~\cite{reagor:2013,romanenko:2020}---also make them a well-suited platform for precision loss measurements.

To further increase sensitivity, one can distinguish dielectric loss from the non-dielectric ``background'' by making a separate measurement of that background, then subtracting it from the loss measured in the presence of the dielectric under study~\cite{checchin:2022}. 
If those measurements are made in separate cooldowns, however, then comparison with the background may be corrupted by cooldown-to-cooldown variation of cavity properties~\cite{turneaure:1968}.
This problem can be avoided by engineering an \textit{in situ} tunable coupling between the cavity-mode and the dielectric under study, which enables both measurements to be performed in the same cooldown. 

\begin{figure}[!tb]
\begin{center}
\includegraphics[width=1\linewidth]{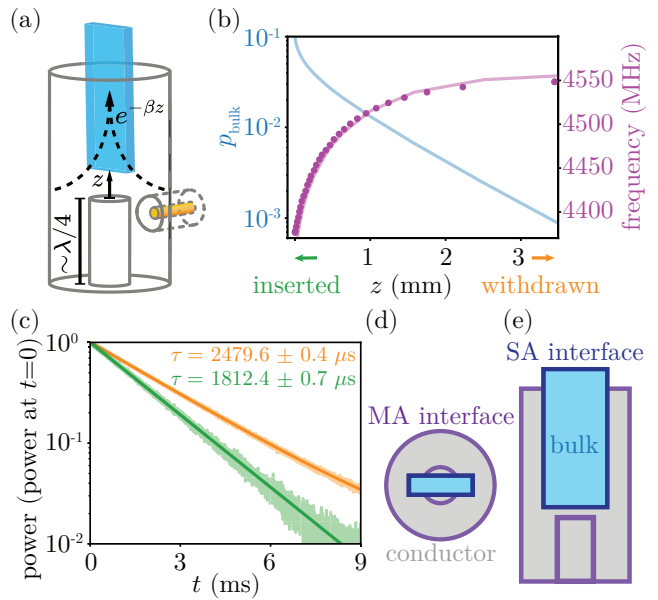}
\caption {\textbf{Measurement concept}. 
(a) A $\lambda/4$ coaxial stub cavity is machined from $4\text{N}$ aluminum, etched~\cite{reagor:thesis}, and mounted to the base of a dilution refrigerator. A dielectric sample is inserted into the cavity waveguide to a position $z$, adjusted \textit{in situ} by a piezoelectric positioner (not pictured).
(b) The fundamental frequency of the cavity is far below the cutoff of the waveguide, making the participation $p_\text{bulk}$ of the cavity mode in the dielectric sample change exponentially with $z$ (blue line).  As $p_\text{bulk}$ rises with insertion, the real part of the sample permittivity causes the measured cavity frequency (purple circles) to drop, in agreement with electromagnetic simulations (purple line).   
(c) Decay of the cavity output power is measured by a ringdown experiment which can distinguish between dephasing and decay (see App.~\ref{app:ringdown}). The imaginary part of the sample permittivity causes the energy decay rate of the cavity mode $1/\tau$ to change when the dielectric is withdrawn (orange) or inserted (green). 
(d) (Top view) In the withdrawn position, almost all of the measured loss comes from either the metal-air dielectric interface (aluminum oxide), or from losses in the conductor.  
(e). (Side view) As the sample is inserted, the loss channels of the substrate under study 
(the substrate-air interface and bulk of the dielectric)
contribute to the measured loss.
}
\label{fig:measurement_cartoon}
\end{center}
\end{figure}

We accomplish this using a coaxial stub cavity~\cite{reagor:2016} with a high-quality (internal quality factor $Q\approx 10^8$) $\lambda/4$ mode, and a piezoelectric positioner that can insert a dielectric sample into the waveguide of the stub cavity (Fig.~\ref{fig:measurement_cartoon}a).
The fundamental cavity mode is designed to have a resonant frequency well below the cutoff of the waveguide. This causes its electric field to be attenuated exponentially along that waveguide, confining the cavity mode.
Because of this attenuation, moving the sample along the waveguide changes the ratio between the electric energy stored in the volume of the dielectric sample $U_\text{bulk}$ and the total energy stored in the cavity $U_\text{tot}$. This ratio is known as the bulk participation $p_\text{bulk}$: 

\begin{eqnarray}
p_\text{bulk}
\equiv
\frac
{U_\text{bulk}}
{U_\text{tot}}
\equiv
\frac
{\frac{1}{2}\int_{V_\text{sub}} \mathbf{D}\cdot \mathbf{E}~dV}
{\frac{1}{2}\int_{V_\text{tot}} \mathbf{D}\cdot \mathbf{E}~dV}
.
\label{p_bulk}
\end{eqnarray}
This ratio $p_\text{bulk}$ can be tuned \textit{in situ} (Fig.~\ref{fig:measurement_cartoon}b), and with a large on-off ratio that is set by the stroke of the piezoelectric positioner and the attenuation length of the waveguide. 

For materials with a permittivity different from vacuum, inserting the dielectric shifts the frequency of the cavity. 
While the positioner does feature a built-in resistive position indication, this frequency shift allows frequency to serve as a proxy for sample position (see Fig. 1b), bypassing the need for the position indication to determine system geometry.
If the permittivity is not purely real, insertion also alters the energy decay rate of the cavity mode in a ringdown experiment, as shown in Fig.~\ref{fig:measurement_cartoon}c (for details of the ringdown measurement, see App.~\ref{app:ringdown}). This \textit{in situ} tunability enables a differential measurement of sample loss to be performed in a single cooldown.

Ideally, all the additional loss observed after insertion of the sample could be attributed to the sample. However, as insertion of the sample perturbs the field distribution of the cavity mode, ``background'' loss from the cavity also changes. This effect can be modeled with the participation formalism, in which the loss of an electromagnetic mode is the weighted-sum of material loss-factors $q_j^{-1}$: 
\begin{eqnarray}
Q^{-1} = \sum_j p_j q_j^{-1},
\label{pf}
\end{eqnarray}
where the weights $p_j \equiv U_j/U_{\text{tot}}$ given to each loss factor are known as the participations~\cite{reagor:thesis} 
(for details on the participations $p_j$, see App.~\ref{app:participations}).  
Implicit in this formalism are the assumptions that mechanisms for loss are linear and uniform within each material region. For comparison with other studies~\cite{wenner:2011,wang:2015}, we assume that surface dielectric interfaces have a thickness of 3 nm (see App.~\ref{app:participations}).
We note that the $q_j^{-1}$ have natural physical interpretations: when $j$ indexes a dielectric loss channel, $q_j^{-1}$ is a loss tangent $\text{tan}\delta_{j}$; and when $j$ indexes a conductor loss channel, $q_j^{-1}$ is the quotient of surface resistance and surface reactance: 
$q_\text{cond}^{-1} = 
{R_\text{s}}/({\omega \mu_0 \lambda_\text{L}})$, 
where $\lambda_\text{L}$ is the London penetration depth~\cite{zmuidzinas:2012}.

We consider a model with losses that originate in the cavity and losses that come from the dielectric sample. Fig.~\ref{fig:measurement_cartoon}d indicates the two material regions of the cavity responsible for ``background'' losses: conductor loss in the superconducting walls of the cavity, and dielectric loss on the metal-air (MA) interface of those walls. While some energy does propagate along the waveguide and radiate out of the cavity, the waveguide length has been chosen to suppress this effect such that $Q^{-1}_\text{radiative} < 10^{-10}$, as verified by electromagnetics simulations. 
This is small compared to the sample losses intended to be resolved in this study, so we neglect radiative loss in our model.

Fig.~\ref{fig:measurement_cartoon}e indicates the two regions associated with loss from the dielectric under study: the substrate-air (SA) interface~\cite{tSA} and the bulk of the dielectric. For a given dielectric sample, these two sources of dielectric loss depend sensitively on the sample position $z$. However, their ratio is predominantly set by the surface-to-volume ratio of the sample, which in the case of a thin sample, is largely insensitive to changes in $z$. This proportionality prevents the separate extraction of these individual loss rates from a measurement of a single sample. For this reason, we combine these two substrate loss channels into a single effective loss channel:

\begin{eqnarray}
\label{eqn:bulk/SA separation}
Q^{-1}_{\text{sub}} \equiv p_{\text{bulk}}q^{-1}_{\text{bulk}} +p_{\text{SA}}q^{-1}_{\text{SA}}.
\end{eqnarray}
From this composite loss channel we define an effective loss tangent of the sample: $q_\text{sub}^{-1}\equiv Q^{-1}_{\text{sub}}/p_{\text{bulk}}$.

Having chosen the loss channels under consideration, a set of measurements $Q^{-1}_{i}$, taken at a series of $N$ different positions $z_i$, can be analyzed using a system of $N$ equations:

\begin{eqnarray}
    \left(\begin{array}{c} 
    Q_{1}^{-1} \\  
    Q_{2}^{-1} \\
    \vdots \\
    Q_{N}^{-1} \\
    \end{array}\right) = \left(\begin{array}{ccc} 
    \tilde{p}_{\text{cond,1}} & p_{\text{MA,1}} & p_\text{{bulk,1}}\\  
    \tilde{p}_\text{{cond,2}} & p_\text{{MA,2}} & p_\text{{bulk,2}}\\  
    \vdots & \vdots & \vdots \\
    \tilde{p}_{\text{cond},N} & p_{\text{MA},N} & p_{\text{bulk},N}\\
    \end{array}\right) \left(\begin{array}{c} 
    \tilde{q}_{\text{cond}}^{-1} \\  
    q_{\text{MA}}^{-1} \\
    q_{\text{sub}}^{-1} \\
    \end{array}\right)
    \label{eqn:Pmat}
    \end{eqnarray}
Here $\tilde{p}_{\text{cond},i}\equiv p_{\text{cond},i}/\omega_i$, and $\tilde{q}_{\text{cond}}^{-1}\equiv q_{\text{cond}}^{-1} \omega_i$. This accounts for the frequency dependence of $q_\text{cond}$.
We refer to the matrix in Eq.~\ref{eqn:Pmat} as the participation matrix $\mathbf{P}$~\cite{reagor:thesis,wang:2015} and determine it with electromagnetic simulations (see App.~\ref{app:participations}).  
The material loss factors $q_{j}^{-1}$ are extracted from Eq.~(\ref{eqn:Pmat}) using bounded (see App.~\ref{app:ps_constraints}) least-squares regression. 

To decompose $q_{\text{sub}}^{-1}$ into its components $q_\text{bulk}^{-1}$ and $q_\text{SA}^{-1}$, a second measurement may be made on a sample of the same material but with different thickness. We demonstrate this process with EFG sapphire in~Sec.~\ref{sec:results}. 

When analyzing various samples measured with the same cavity, we concatenate multiple instances of Eq.~(\ref{eqn:Pmat}), providing each sample its own substrate loss factor and associated matrix column.
Because this joint analysis permits only a single value of $q_{\text{MA}}^{-1}$ and $q_{\text{cond}}^{-1}$, the number of free parameters in the model is reduced by four (compared to fitting three samples individually), ensuring consistency in the extracted substrate loss factors.
Joint analysis is valid only if the cavity's bare losses do not change between cooldowns but this can be verified by measuring the cavity with each sample in the withdrawn position.

\subsection{Sensitivity}

When solving for the inferred material loss factors $q_j^{-1}$, the structure of $\mathbf{P}$ determines how errors in the measured $Q_i^{-1}$ propagate into loss factor errors $\sigma_{q_j^{-1}}$. If the propagated fractional error $\sigma_{q_j^{-1}}/q_j^{-1}$ is greater than 100\%, then the loss factor is said to be unresolved as it is below the sensitivity of the measurement.
One option for predicting $\sigma_{q_j^{-1}}$ is to use the condition number~\cite{trefethen:1997} of $\mathbf{P}$ to infer a bound on the severity of this error amplification. Such a perspective is useful in experimental design, and has been used to improve the conditioning of matrix problems similar to Eq.~(\ref{eqn:Pmat})~\cite{calusine:2018,woods:2019}.

Another tool for predicting the error and sensitivity of an experiment is the covariance matrix. Given some participation matrix $\mathbf{P}$ and an approximate expectation for the measurement errors $\sigma_{Q_i^{-1}}$, the covariance matrix $\mathbf{C}$ allows the individual $\sigma_{q_j^{-1}}$ to be calculated (and not just bounded) from its diagonal elements $C_{jj} = \sigma_{q_j^{-1}}^2$.  For the case of linear least-squares, 
\begin{eqnarray}
\mathbf{C} \equiv (\mathbf{\tilde{P}}^\text{T}\mathbf{\tilde{P}})^{-1},
\label{cov}
\end{eqnarray}
where $\tilde{P_{ij}} \equiv P_{ij}/\sigma_{Q_i^{-1}}$~\cite{richter:1995}.
The specificity of this information makes the covariance matrix a powerful tool for experimental design. We use it to predict the sensitivity of the dielectric dipper. 

\begin{figure}[!t]
\begin{center}
\includegraphics[width=1\linewidth]{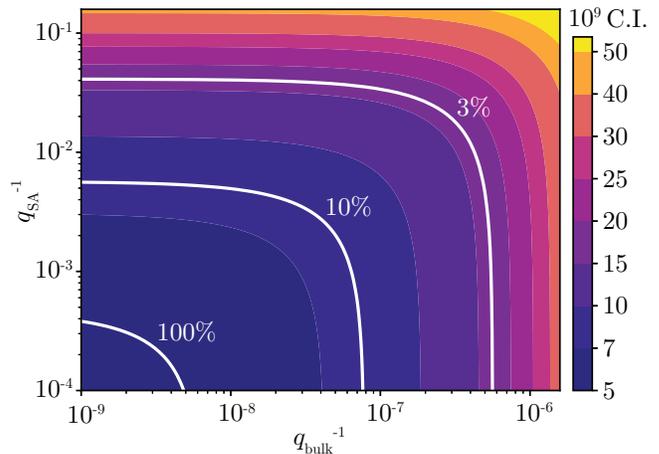}
\caption{\textbf{Simulated sensitivity}. Estimated $q_\text{sub}^{-1}$ 95\% confidence interval ($\text{C.I.}$), plotted (in color) as a function of hypothetical dielectric material quality factors $q_\text{SA}^{-1}$ and $q_\text{bulk}^{-1}$.  
White curves indicate the $3\%$, $10\%$ and $100\%$ fractional error contours.  We assume fractional measurement error $\sigma_{Q_i^{-1}}/Q_i^{-1} = 1\%$ (limited by our ability to separate coupling loss from total loss), cavity material quality factors $q_\text{MA}^{-1} = 3\times10^{-2}$, and $q_\text{cond}^{-1} = 2\times10^{-5}$ and $N=30$ measurement positions.
Participations are as simulated for the 460\micron~thick EFG sapphire sample.
}
\label{fig:sensitivity}
\end{center}
\end{figure}

Accurate prediction of an extracted loss factor error $\sigma_{q_j^{-1}}$ by solving Eq.~\ref{cov} requires two things: the participation matrix $\mathbf{P}$ (obtained from simulations) and an anticipated absolute measurement error, $\sigma_{Q_i}^{-1}$. 
The error in the measurement of a line-width often scales with the line-width, so we assume a fractional measurement error $\sigma_{Q_i^{-1}}/Q_i^{-1}$. The conversion an anticipated fractional error into an anticipated absolute error requires an accurate prediction of $Q_{i}^{-1}$, which can be made using participations from simulation, and assumed values for $q_j^{-1}$ as inputs to eq.~(\ref{pf}). In this experiment, reasonable values for $q_\text{cond}^{-1}$ and $q_\text{MA}^{-1}$ can be chosen during the design phase from measurements of a cavity made of the same materials as the planned design (or the exact cavity to be used if it is available). However, during the design phase, the substrate material loss factors $q_\text{bulk}^{-1}$ and $q_\text{SA}^{-1}$ are unknown.  We therefore calculate $\sigma_{q_\text{sub}^{-1}}$ as a function of these unknown substrate loss factors. The result of this error analysis is plotted in color in Fig.~\ref{fig:sensitivity} in terms of the 95\% confidence interval $\text{C.I.} = 2 \sigma_{q_\text{sub}^{-1}}$. Also seen in Fig.~\ref{fig:sensitivity} are white contours which indicate the result of converting these absolute errors into fractional errors, $2\sigma_{q_{\text{sub}}^{-1}}/q_{\text{sub}}^{-1}$. These provide a visualization of which loss tangent combinations would be resolvable by this measurement with 95\% confidence.

When the losses from the sample are sufficiently small, the fractional error exceeds one and the substrate material quality factor $q_\text{sub}^{-1}$ can no longer be distinguished from zero. The $100\%$ fractional error contour indicates the boundary of this region. The calculated error near the boundary of this region, $\sigma_{q_\text{sub}^{-1}}<5$ parts-per-billion (ppb), indicates the predicted sensitivity of the experiment. The sensitivity of our experiment is predominantly limited by our ability to know the composition of the cavity losses, and in turn our ability to distinguish substrate losses from losses in the metal-air interface of the cavity (for details, see App.~\ref{app:poly_sens}). 
Thus, even though multiple participations change with sample insertion to some degree, we can still measure dielectric losses with 5 ppb sensitivity.

\section{Sapphire Measurements}
\label{sec:results}

\begin{figure*}[tbh]
\begin{center}
\includegraphics[width=1\linewidth]{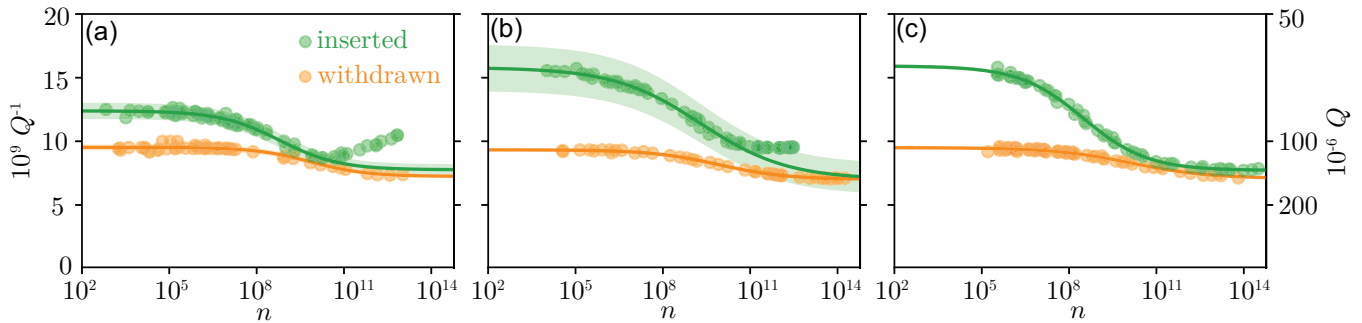}
\caption {\textbf{Power dependence of loss}. Measured internal loss $Q^{-1}$ with the sample withdrawn (orange) and inserted (green) as a function of the average initial cavity photon number $n$ in the ringdown measurement, with (a)~$100$\micron~thick EFG sapphire, (b)~$460$\micron~thick EFG sapphire, and (c)~$440$\micron~thick HEMEX sapphire. 
The curves are fits of the TLS model in Eq.~(\ref{eqn:TLS}) and the shaded region reflects propagation of the standard error of the fit parameters $q_{j}^{-1}$ to $Q^{-1}$. 
For the inserted measurements of EFG sapphire, fits are only to measurements with $n \lesssim 2\times10^{10}$, as above this photon number the loss deviates from the TLS model. The origin of this deviation is unclear but we conjecture it to be related to heating of the dielectric sample.
}
\label{fig:pow_sweep}
\end{center}
\end{figure*}

To demonstrate the utility of the dielectric dipper, we use it to determine whether bulk dielectric loss plays a role in the lifetimes of state-of-the-art superconducting qubits on sapphire~\cite{place:2021,wang:2022}. 
To do this, we measure samples from wafers of sapphire cut from sheets grown by the same edge-defined film-fed growth method (EFG)~\cite{harris:2003} as the substrates used in those studies. Samples of two different thicknesses (and the same length and width) were measured in order to localize substrate loss to either the bulk or the substrate-air interface using Eq.~(\ref{eqn:bulk/SA separation}). Motivated by previous studies~\cite{creedon:2011,dial:2016} on low-loss sapphire substrates grown with a different process, we also measure a third sapphire sample which is cut from a boule grown by the heat-exchanger method (HEM)~\cite{harris:2003}. After production, this wafer was screened by the manufacturer for low optical absorption and wavefront aberration, qualifying for the highest grade: HEMEX~\cite{khattak:1997}. 
For all three samples, we perform power sweeps with the sample inserted to ensure that following measurements are made in the low power regime. Withdrawn power sweeps are also performed to infer bounds on the background losses of the cavity. 
We then perform a position sweep at low excitation power to extract the substrate loss tangents.

\subsection{Power-dependence of loss}
Due to the saturability of dielectric loss, measurements which aim to be representative of qubit performance must be taken in this same low-power regime where superconducting qubits operate.
To identify this regime, we begin with the sample completely inserted, and measure the cavity's impulse response as a function of the excitation power in the ringdown experiment. 
We parameterize cavity excitation by the average number of photons $n$ created by the impulse (see App.~\ref{sec:nbar}). The green circles in Fig.~\ref{fig:pow_sweep} show the results for the 100\micron~EFG sample (a), the 460\micron~EFG sample (b), and the 440\micron~HEMEX sample (c).  In each case, 
the total internal loss of the cavity asymptotically approaches some high-power value $Q^{-1}_\text{hp}$, and some low-power value $Q^{-1}_\text{lp} = Q^{-1}_\text{hp} + Q^{-1}_\text{sat}$, with a smooth transition between these two regimes. 
This power-dependence can be captured by a two-level-system (TLS) model for dielectric loss~\cite{muller:2019} which we fit to the data:

\begin{eqnarray}
Q^{-1} = Q^{-1}_\text{hp} + \frac{Q^{-1}_\text{sat}}{\sqrt{1+\left(\frac{n}{n_\text{c}}\right)^\alpha}}.
\label{eqn:TLS}
\end{eqnarray}
Here the critical photon number $n_\text{c}$ sets the scale for saturation and $\alpha$ sets the width of this transition. Table~\ref{tab:pow_sweep} shows the extracted fit parameters for these sample-inserted power sweeps.

\begin{table}[htb]
\begin{tabular}{ |c||c|c|c| } 
 \hline
 Sample & $n_\text{c}$ (millions)& $\alpha$ \\ \hline
 $100$\micron~EFG sapphire & $600(300)$ &  $0.42(5)$\\ 
 $460$\micron~EFG sapphire & $2000(1800)$ & $0.29(4)$\\
 $440$\micron~HEMEX sapphire  & $300(30)$ & $0.40(1)$\\ \hline
\end{tabular}
\caption {Fit results for the inserted power sweeps in Fig.~(\ref{fig:pow_sweep}) using the TLS model in Eq.~(\ref{eqn:TLS}).} 
\label{tab:pow_sweep}
\end{table}

The inserted power sweeps show that cavity population $n=10^5$ is sufficiently below $n_\text{c}$, such that measured loss is only weakly dependent on power and closely approximates the low-power behavior $Q^{-1}_\text{lp}$, as is appropriate for predicting their performance as superconducting qubit substrates.
We therefore fix the excitation photon number for the remaining measurements to $n=10^5$ (for details, see App.~\ref{app:lp_regime}). This results in a peak electric field in the sample of about $4 \times 10^{-8}$~V/m.  
Note that the extracted values of $n_\text{c}$ are a factor of $10^6$ greater than those found in coplanar waveguide resonators~\cite{quintana:2014,dunsworth:2017}, reflecting the much larger mode volume of a three-dimensional cavity~\cite{lei:2020}. 

To measure the power dependence of the cavity background alone and bound the low-power values of $q_\text{cond}$ and $q_\text{MA}$, we repeat the power sweep with the sample completely withdrawn (orange circles in Fig.~\ref{fig:pow_sweep}).
Previous studies on conductor losses in aluminum cavities have found no power dependence up to $10^{10}$ photons~\cite{reagor:2013}.  If we assume that this trend continues up to $10^{14}~\text{photons}$, the measured power dependence may be attributed entirely to saturation of dielectric loss from the metal-air interface of the cavity.  This measurement therefore provides two useful constraints on the cavity loss factors $q_\text{cond}^{-1}$ and $q_\text{MA}^{-1}$: first, the conductor loss factor is bounded from above by the measurement with the smallest loss $\text{min}(Q^{-1})$ (typically a high-power measurement, ideally described by $Q_\text{hp}^{-1}$);  and second, the metal-air loss factor is bounded from below by the change in the cavity loss $\Delta Q^{-1}$ (ideally described by $Q_\text{sat}^{-1}$).  More precisely,
\begin{eqnarray}
\label{eq:psweep_bounds}
q_\text{cond}^{-1} &\leq& \frac{\text{min}(Q^{-1})}{p_\text{cond}}, \\ \nonumber
q_\text{MA}^{-1} &\geq& \frac{\Delta Q^{-1}}{p_\text{MA}}.
\end{eqnarray}
Table~\ref{tab:cav_loss_bounds} in App.~\ref{app:ps_constraints} lists the bounds obtained in this way.

\begin{figure*}[!t]
\begin{center}
\includegraphics[width=1\linewidth]{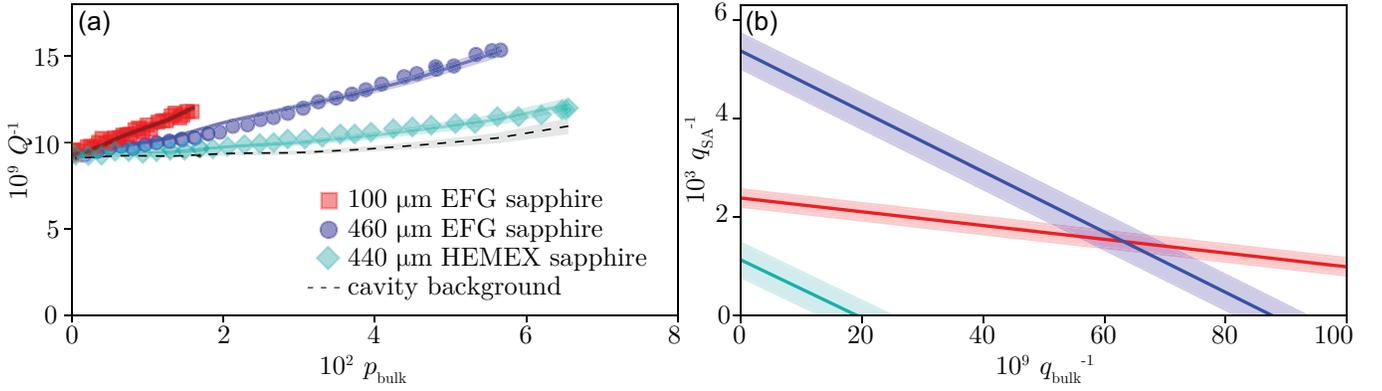}
\caption {\textbf{Measurements of the bulk and surface loss tangents}. (a) Measured internal loss $Q^{-1}$ plotted as a function of bulk participation $p_\text{bulk}$, for $100$\micron~thick EFG sapphire, $460$\micron~thick EFG sapphire, and $440$\micron~thick HEMEX sapphire. The excitation pulse for these measurements create approximately $10^5$ photons (low-power regime), and the mixing chamber temperature is 20 mK (low-temperature regime).  Solid lines are fits to the three-loss model in Eq.~(\ref{eqn:Pmat}), consistent with bounds provided by the power sweeps (see App.~\ref{app:ps_constraints}). The shaded region shows propagation of the 95\% confidence interval of $q^{-1}_\text{sub}$, and the dashed line is the estimated cavity background, comprised of conductor and MA dielectric losses. (b) The effective substrate loss factor $q_\text{sub}^{-1}$ is a weighted average of substrate bulk and surface loss factors.  By Eq.~(\ref{eqn:bulk/SA separation}), a measurement of $q_\text{sub}^{-1}$ defines a line in the two-dimensional parameter space of $q_\text{bulk}^{-1}$ and $q_\text{SA}^{-1}$.   
The intersection of two such lines, each defined by measurements on the same material with different form factors, identifies the bulk and SA loss factors of that material.
The single measurement of HEMEX sapphire does not separately resolve bulk and SA losses, but bounds both the bulk and SA loss factors.
}
\label{fig:insertion_sweep}
\end{center}
\end{figure*}

\subsection{\textit{In situ} insertion of the sample}

With the cavity loss factors constrained and the edge of the low-power regime identified, we proceed with the position sweeps described in Sec.~\ref{sec:Measurement technique}.  The results are shown in  Fig.~\ref{fig:insertion_sweep}a. For each sample in this study, internal loss $Q^{-1}$ of the cavity mode is plotted against the substrate bulk participation $p_\text{bulk}$. When the substrate is withdrawn from the cavity, $p_\text{bulk}$ approaches zero, and the measured loss is that of the empty cavity. As the sample is inserted, $p_\text{bulk}$ rises, and the cavity mode inherits loss from the sample. All samples are inserted until $z\approx 30$\micron.

Examining the position sweeps for the two EFG samples, one can see that the total change in $Q^{-1}$ is larger for the thicker sample.  As the samples have almost identical surface areas, this indicates that surface loss does not account for all of the added loss.
To check that intuition and distinguish the bulk and surface contributions, we first extract the substrate loss tangent $q_\text{sub}^{-1}$ for each sample by solving Eq.~(\ref{eqn:Pmat}) (constrained by Eq.~(\ref{eq:psweep_bounds})).  This separates substrate loss from the ``background'' cavity losses.  As $q_\text{sub}^{-1}$ is a weighted sum of the bulk and surface loss tangents of the sample material per Eq.~(\ref{eqn:bulk/SA separation}), a measurement of $q_\text{sub}^{-1}$ constrains those loss tangents, in the space of all possible ordered pairs $(q_\text{bulk}^{-1},q_\text{SA}^{-1})$, to lie along a line (see Fig.~\ref{fig:insertion_sweep}b).  A single constraining line has several useful geometric interpretations: its $x$ and $y$ intercepts correspond to upper bounds on the bulk and SA interface loss tangents; its distance from the origin is proportional to the measured value of $q_\text{sub}^{-1}$; and its slope is determined by the ratio of surface and bulk participations $p_{\text{SA}}/p_{\text{bulk}}$.  

Because the two EFG sapphire samples have different surface-to-volume ratios, the constraint lines associated with these two measurements have different slopes.  Their intersection marks the only viable $(q_\text{bulk}^{-1}$ , $q_\text{SA}^{-1})$ pair which is consistent with the measured $q_\text{sub}^{-1}$ of both samples.
In this way, we distinguish between bulk loss and surface loss in EFG sapphire and extract $q_\text{bulk}^{-1} = 63(8)\times10^{-9}$, and $q_\text{SA}^{-1} = 15(3)\times10^{-4}$~\cite{tSA}. 

Fig.~\ref{fig:insertion_sweep}a also displays the results of the position sweep on a HEMEX sapphire sample. The more gradual slope of this data set, compared to the EFG data sets, indicates that HEMEX sapphire has lower dielectric losses.  
Having measured one HEMEX sample, there is only a single contour to constrain the possible $(q_\text{bulk}^{-1}$ , $q_\text{SA}^{-1})$ pairs, so the bulk and surface contributions are not separately resolved.  Nonetheless, the measurement bounds both bulk and surface losses. These bounds are represented graphically by the $x$ and $y$ intercepts of the HEMEX constraint line in  Fig~\ref{fig:insertion_sweep}b.
The inferred bound on bulk loss is $q_\text{bulk}^{-1} < 19(6)\times10^{-9}$, consistent with the findings of Ref.~\cite{creedon:2011}, and smaller than the bulk loss tangent of EFG sapphire by a factor of 3 (see Table~\ref{tab:qbulk}.)

\begin{table}[htb]
\begin{tabular}{ |c||c|c| } 
 \hline
 Material       & $q^{-1}_\text{bulk}$ & $q^{-1}_\text{SA}$\\ \hline
 EFG sapphire   & $63(8)\times10^{-9}$                       & $15(3)\times10^{-4}$ \\ 
 HEMEX sapphire & $<19(6)\times10^{-9}$                         & $<11(4)\times10^{-4}$   \\ \hline
\end{tabular}
\caption {Measurements and bounds of the bulk and surface dielectric loss tangents for EFG and HEMEX sapphire~\cite{tSA}.}
\label{tab:qbulk}
\end{table}

\section{Implications for transmon lifetimes} 
\begin{figure}[!htb]
\begin{center}
\includegraphics[width=1\linewidth]{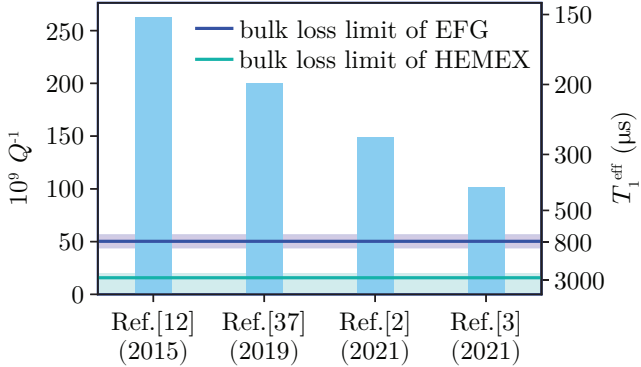}
\caption {\textbf{Inferred bulk coherence limit and comparison to measured transmon lifetimes}.  The coherence limit imposed by EFG sapphire and HEMEX sapphire assuming a bulk participation of $p_\text{bulk} = 0.8$, compared to quality factors of transmons on sapphire from select studies. 
Comparisons are made in terms of total loss $Q^{-1}$ and in terms of $T_{1}^{\text{eff}}$; a $T_{1}$ calculated from $Q^{-1}$ assuming a $4$ GHz qubit transition frequency.}
\label{fig:transmon_implications}
\end{center}
\end{figure}

Having sensitively measured the loss tangents of these forms of sapphire, we now calculate the current and future coherence limits that they imply for typical transmon qubits.
Assuming  a qubit transition frequency of $4$ GHz and a bulk participation of $p_\text{bulk}= 80\%$, these measurements imply that a transmon fabricated on EFG sapphire would be bulk limited to a quality factor of $Q\leq 20 \times 10^6$, or $T_1 \leq 800~\upmu$s.
Several comments can be made on this coherence limit, which is illustrated in Fig.~\ref{fig:transmon_implications} alongside the coherence reported in several recent studies of transmons on sapphire.

First, this limit is consistent with previous studies~\cite{wang:2015} in which transmon lifetimes were found to be limited by surface losses, as those studies featured devices with quality factors up to $4 \times 10^6$, and therefore would not have observed a strong effect from a coherence limit at the level of $20 \times 10^6$. This is illustrated in Fig.~\ref{fig:transmon_implications} by the EFG sapphire bulk loss limit being significantly below the total loss rates $Q^{-1}$ observed in that study.

Second, this limit is consistent with the best-reported lifetimes of transmons made on EFG sapphire~\cite{serniak:2019,place:2021,wang:2022}, as no such study has yet produced a transmon with quality factor $Q>20 \times 10^6$. Interestingly, the quality factors of the longest lived-transmons are approaching this bound, as seen in Fig.~\ref{fig:transmon_implications}. Supposing that the fabrication process used to make those transmons does nothing to improve the bulk loss tangent of the substrate, this comparison suggests those transmons may be significantly bulk limited. 
For bulk-limited devices, replacing EFG with HEMEX sapphire will substantially increase transmon lifetimes. The increased coherence limit is also shown in Fig.~\ref{fig:transmon_implications}.  Further, the shared chemistry and lattice constant of these materials implies that this substitution would not require any  modification of the fabrication process. 

A general insight gained from this comparative measurement is that not all sapphire is equal.  As with high-power measurements made at several kelvin~\cite{strayer:1983,braginsky:1987,luiten:1993,krupka:1999b}, growth method has a meaningful impact on the millikelvin-and-low-power loss tangent of sapphire, and in turn the lifetime of transmons built upon that sapphire.

Additionally, for completeness we use the measured value of $q_\text{SA}^{-1}$ to calculate the SA coherence limit for a typical transmon. Assuming $p_\text{SA}=1.18\times10^{-4}$~\cite{wang:2015},  substrate-air surface dielectric loss would limit the quality factor of a transmon to $Q\leq 7 \times 10^6$.  In real devices, however, the loss tangent $q_\text{SA}^{-1}$ will likely depend on the details of the substrate surface preparation.  More definite claims can be made about the SA-interface coherence limit of real devices by leveraging the flexibility of the dipper technique and measuring dielectric samples that have undergone the same surface preparation sequence as the substrates used for those devices
(for details on the preparation of samples in this study, see App.~\ref{app:provenance}).

Finally, the model used in our analysis has assumed that all sample loss is dielectric in origin rather than magnetic, but we can extend the model and allow sample loss to be a combination of dielectric and magnetic contributions. By considering our results for EFG sapphire in comparison to performance of striplines from a previous study~\cite{axline:2016}, we can make two bounding statements about magnetic bulk loss in EFG sapphire. First, the EFG bulk loss tangent measured in this study must be at least 97\% truly dielectric in origin, and no more than 3\% could be from misattributed magnetic loss. This validates the choice to neglect magnetic sample loss in our analysis. Second, bulk magnetic loss is not a major source of loss in even the longest-lived transmons on EFG sapphire, accounting for less than 10\% of the total loss. (for details, see App.~\ref{app:mag_loss}).

\section{Conclusion}

Precise knowledge of the decoherence mechanisms in superconducting qubits aids in all efforts to improve their coherence. Here, we present a method to precisely measure bulk dielectric loss with $5\times10^{-9}$ sensitivity, and apply it in a study of EFG sapphire.  At low powers and temperatures, we measure the bulk loss tangent at microwave frequencies to be $63(8) \times 10^{-9}$.  This entails a bulk limit to the quality factor of a typical transmon on an EFG sapphire substrate of $Q \lesssim 20 \times 10^6$, consistent with the longest lifetimes of transmons on sapphire, and suggests bulk dielectric loss is a major source of loss in those devices.  We also measure HEMEX sapphire, and bound its bulk loss tangent to be less than $19(6)\times10^{-9}$.  Such a low-loss substrate would reduce one of the dominant sources of loss by a factor of 3 or more, and could enable longer transmon lifetimes.

Looking forward, measurements of other materials, such as silicon, quartz, or lithium niobate, can inform efforts to improve coherence of a variety of devices, including qubits, on those substrates.  We also expect that correlation of precision dielectric loss measurements with other physical probes (e.g. x-ray spectroscopy or scattering) will shed light on the underlying physical mechanisms behind this loss.  

\vspace{0.1in}
We acknowledge N. Ofek, Y. Liu, and  P. Reinhold, for their work building the FPGA firmware and software used in this experiment, C. J. Axline and K. Li for their contributions in developing earlier versions of the experiment, and N. P. de Leon, A. Walter, A. Barbour, M. H. Devoret and S. M. Girvin for useful discussions.  Use of facilities was supported by the Yale Institute for Nanoscience and Quantum Engineering (YINQE), Yong Sun, and the Yale SEAS cleanroom.  We thank MIT Lincoln Laboratory and the Intelligence Advanced Research Projects Activity (IARPA) for providing a Josephson travelling wave parametric amplifier.  This research, including funding for A.P.R., was supported by the U.S. Department of Energy, Office of Science, National Quantum Information Science Research Centers, the Co-design Center for Quantum Advantage (C2QA) under contract number DE-SC0012704. Further support, including funding for B.J.C., came from the Army Research Office (ARO), under Grant Number W911NF-18-1-0212. The views and conclusions contained in this document are those of the authors and should not be interpreted as representing the official parties, either expressed or implied, of the Army Research Office (ARO) or the U.S. Government. The U.S. Government is authorized to reproduce and distribute reprints for government purposes notwithstanding any copyright notation herein.

B.J.C. and R.J.S. conceived the experiment; A.P.R. and B.J.C. designed and characterized the experimental apparatus, carried out the experiments, analyzed the data, and wrote the manuscript with input from all authors; L.F. and R.J.S. supervised the work.

L.F. and R.J.S. are founders and shareholders of Quantum Circuits Inc.
\\

\appendix
\section{Ringdown measurement}
\label{app:ringdown}

\begin{figure}[!htb] 
\begin{center}
\includegraphics[width=1\linewidth]{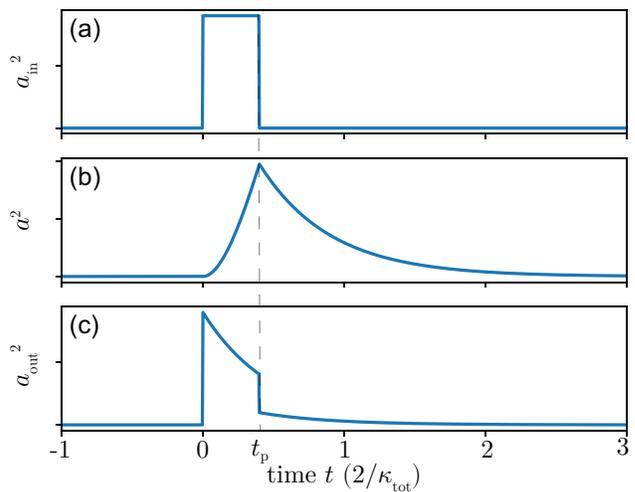}
\caption {\textbf{Dynamics of a resonant ringdown measurement}: (a)~the incoming field $a_\text{in}$, (b)~cavity mode amplitude $a$, and (c)~the outgoing field $a_\text{out}$, each as a function of time.  Figure is not to scale.
} 

\label{fig:ringdown_cartoon}
\end{center}
\end{figure}

To measure the internal energy decay rate $\kappa_{\text{int}}$ of a microwave cavity with (classical) mode amplitude $a$, the cavity is excited with an input field $a_\text{in}$ (Fig.~\ref{fig:ringdown_cartoon}a) and the output field $a_\text{out}$ is monitored to determine the rate at which its energy is lost.  
 
During the pulse, the cavity mode amplitude $a$ rises. After the pulse, $a$ decays exponentially (Fig.~\ref{fig:ringdown_cartoon}b).
While a portion of the energy from the excitation pulse makes its way into the cavity mode, the rest is reflected off the cavity port and propagates to the receiver (for the complete wiring diagram, see App.~\ref{app:circuit}). At the same time, the cavity emits a field proportional to $a$. The reflected pulse and the cavity emission interfere to create the output field $a_\text{out}$, which is measured at the receiver (Fig.~\ref{fig:ringdown_cartoon}c). 

We determine the total loss of the cavity mode $\kappa_{\text{tot}}$ by fitting the decay rate of $|a_\text{\text{out}} |^2$ for times $t>t_\text{p}$, during which $a_{\text{out}} \propto a$. 
To distinguish between the effects of dephasing and actual energy decay, the emitted field must be sampled at a rate faster than the dephasing time of the cavity (see App.~\ref{app:bandwidth}), and data from multiple shots of the experiment must be processed with a phase-insensitive averaging scheme (see App.~\ref{app:dephasing}).

The external coupling-rate of the cavity mode $\kappa_{\text{ext}}$ is encoded in the overall amplitude of the emitted field in comparison to the amplitude and duration of the excitation pulse and can be inferred from the outgoing field after the beginning ($t = 0^+$) and after the end ($t = t_\text{p}^+$) of the excitation pulse (see App.~\ref{app:coupling}).

Those two loss rates can then be subtracted to determine the internal energy decay rate $\kappa_{\text{int}} = \kappa_{\text{tot}} - \kappa_{\text{ext}}$, 
which can be divided by the resonant frequency $\omega_a$ of the cavity to yield a dimensionless loss: $Q^{-1}_{\text{int}} = \kappa_{\text{int}}/\omega_\text{a}$.

\subsection{Decay of a harmonic oscillator with time-varying resonant frequency}
\label{app:dephasing}
\begin{figure}[!htb] 
\begin{center}
\includegraphics[width=1\linewidth]{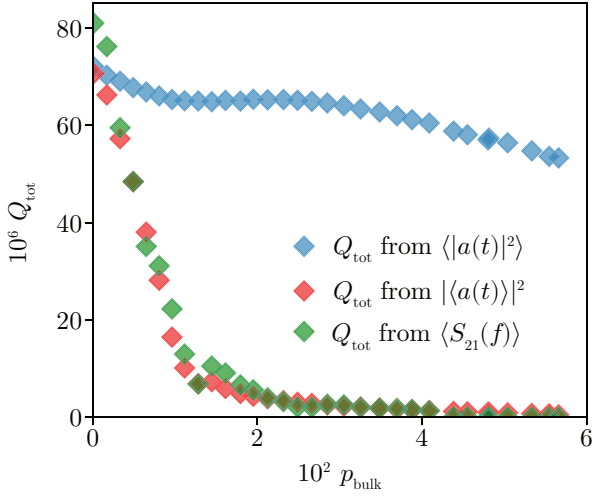}
\caption {\textbf{Consistency between ringdown and spectroscopic measurement}. When the sample is withdrawn from the cavity, $Q_\text{tot}$ inferred from the decay rates of $\langle |a(t)|^2 \rangle$ (blue) and $|\langle a(t) \rangle|^2$ (red) are in close agreement with each other, and with $Q_\text{tot}$ from circle fits~\cite{petersan:1998} of measurements taken by a vector network analyzer $\langle S_{21}(f) \rangle$ (green). As the sample is inserted, energy decay and frequency jitter tend to reduce $Q_\text{tot}$ extracted from $|\langle a(t) \rangle|^2$ or $\langle S_{21}(f) \rangle$, however $\langle |a(t)|^2 \rangle$ is not susceptible to frequency jitter (see Eq.~\ref{eqn:T1}) and thus only experiences reduction in $Q_\text{tot}$ caused by energy decay. $Q_\text{tot}$ can be, in general, non-monotonic in $p_\text{bulk}$ due to frequency-dependent coupling loss, which is measured and subtracted off before further processing (see App.~\ref{app:coupling}).
} 
\label{fig:T1vsT2vsVNA}
\end{center}
\end{figure}

In the presence of a fluctuating cavity frequency, the cavity output field $|\langle a_{\text{out}} \rangle|$ (averaged over many shots) does not decay at the energy decay rate $\kappa_\text{tot}$. Rather, it decays at a rate which is also affected by cavity dephasing. For this reason, measurements of $|\langle a_{\text{out}} \rangle|$, such as those from a vector network analyzer, cannot be used to reliably extract $\kappa_\text{tot}$. Instead, one must measure the ensemble-average of the cavity output power, $\langle |a_\text{\text{out}}| ^2 \rangle$, which is a phase-insensitive quantity and decays at the same rate, $\kappa_\text{tot}$, as the cavity energy. In many cases, this distinction is irrelevant, but in our experiment, mechanical vibrations of the high-permittivity dielectric sample cause the cavity frequency to jitter by as many as 100 line widths (see Fig.~\ref{fig:T1vsT2vsVNA}). When the dielectric is inserted, accurate measurement of $\kappa_\text{tot}$ therefore requires fitting the decay of $\langle |a_\text{out} |^2 \rangle$, rather than that of $|\langle a_\text{out} \rangle|^2$. In this section we describe how the energy decay rate $\kappa_\text{tot}$ can be inferred from ringdown measurements of an oscillator with a fluctuating resonance frequency.

Suppose an oscillator has a time-varying resonant frequency
~$\omega_\text{a} = \omega_0(1 + \epsilon(t))$.
Assume the fluctuations of the resonance $\epsilon(t)$ are caused by a stochastic process with zero mean,
\begin{eqnarray}
\langle\epsilon(t)\rangle &=& 0, 
\label{zeromean}
\end{eqnarray}
and characterized by a power spectral density $S(f)$:
\begin{eqnarray}
\langle \epsilon(t)^2\rangle &=& \int_0^\infty df S(f) \cos(2 \pi f t).
\end{eqnarray}
where the brackets $\langle\cdot\rangle$ denote ensemble averaging.

The Langevin equation~\cite{collet:1983} dictates how $a$ evolves in time. In the absence of an excitation pulse (here, true for $t > t_\text{p}$), the measured output field $a_\text{out}$ is proportional to $a$ and governed by an equation of the same form:
\begin{eqnarray}
\dot{a}_\text{out} 
&=& 
[i \omega_0 (1 + \epsilon(t)) - \kappa_\text{tot}/2] a_\text{out}.
\label{eqn:adot}
\end{eqnarray}
After separation of variables, integration yields
\begin{eqnarray}
a_\text{out}(t) 
&=& 
e^{(i[\omega_0 t + \theta(t)] - \kappa_\text{tot} t/2)},
\end{eqnarray}
with
\begin{eqnarray}
\theta(t) &\equiv& \omega_0 \int_0^t \epsilon(t')dt'.
\end{eqnarray}
Here we have chosen the constant of integration to reflect an initial condition of unit amplitude in the oscillator mode.

The average of the output power decays as
\begin{eqnarray}
\langle |a_\text{out}(t)|^2\rangle = e^{-\kappa_\text{tot} t}.
\label{eqn:T1}
\end{eqnarray}
We identify the decay time of the oscillator as $\tau \equiv 1/\kappa_\text{tot}$.
$\langle |a_\text{out}(t)|^2\rangle$ is independent of $\theta(t)$ and is thus unaffected by the fluctuations in the resonant frequency. 

This is in contrast to the average of the output field,
\begin{eqnarray}
|\langle a_\text{out}(t)\rangle| &=& e^{-\kappa_\text{tot} t/2} |\langle e^{i \theta(t)}\rangle|, \\ \nonumber
&=& e^{-\kappa_\text{tot} t/2} \left(1 - \frac{1}{2} \langle \theta(t)^2\rangle +\mathcal{O}(\theta(t)^3)\right).
\end{eqnarray}
By our assumption that the fluctuations are zero-mean, the linear term in the expansion vanishes.
(If the noise is Gaussian, the higher moments vanish and this approximation is exact.) To evaluate the quadratic term, we follow the treatment given in Ref.~\cite{martinis:2003}:
\begin{eqnarray}
\langle\theta(t)^2\rangle 
&=& \omega_0^2 \int_0^t\int_0^tdt'dt'' \langle\epsilon(t')\epsilon(t'')\rangle, 
\\ \nonumber
&=& \omega_0^2 \int_0^t\int_0^tdt'dt'' \int_0^\infty df S(f) \textrm{Re}\{ e^{(2 \pi i f (t'-t''))}\}, 
\\ \nonumber
&=& \omega_0^2 \int_0^\infty df S(f) W_0(f,t),
\label{thetas}
\end{eqnarray}
where the spectral weight function $W_0(f,t)$
\begin{eqnarray}
W_0(f,t) &=& \left|\int_0^t dt' e^{2\pi i f t'}\right|^2, \\ \nonumber
&=& \frac{\sin(\pi f t)^2}{(\pi f)^2},
\end{eqnarray}
low-pass filters the fluctuations.

We therefore have
\begin{eqnarray}
\label{eqn:T2fast}
|\langle a_\text{out}\rangle| &=& e^{-\kappa_\text{tot}t/2} \times  \\ 
&& \left(1 - \frac{1}{2} \omega_0^2 \int_0^\infty df S(f) W_0(f,t) \nonumber +\mathcal{O}(\langle\theta(t)^3\rangle)\right).
\end{eqnarray}
The exponential prefactor represents energy decay from the oscillator and the rest of the expression accounts for the effect of pure dephasing. The form of this dephasing depends on the noise spectrum $S(f)$.

In summary, the exponential decay of the phase-insensitive quantity $\langle |a_\text{out}|^2\rangle$ yields the time-scale for energy decay $\tau$, and the phase-sensitive quantity $|\langle a_\text{out}\rangle|$ has a $1/e$ time which is affected by both energy decay and dephasing in the cavity. 
This is analogous to the distinction between $T_1$ and $T_2$ of a spin ensemble. Fluctuations in the cavity's resonant frequency can be detected by comparing $\tau$ to the $1/e$ time of $|\langle a_\text{out}\rangle|^2$. By converting these decay rates to quality factors, they can be checked for consistency with quality factors extracted from circle fits of spectroscopic measurements~\cite{petersan:1998} (See Fig.~\ref{fig:T1vsT2vsVNA}).

\subsection{Finite measurement bandwidth}
\label{app:bandwidth}
The above arguments can be generalized to account for the effects of finite measurement bandwidth, such as that imposed by a non-instantaneous measurement. In short, finite measurement bandwidth causes a reduction in the contrast of the measurement $\langle |a^\text{m}_\textrm{out}|^2\rangle$, but does not change the time scale for decay.  For completeness, we present those arguments below, as the suppression of contrast is an effect that must be controlled in order to accurately measure $Q_\text{ext}^{-1}$ (a measurement discussed in the next subsection,  App.~\ref{app:coupling}).

The result of a finite bandwidth measurement of the output field $a^\text{m}_\text{out}(t)$ is the convolution of the signal $a_\text{out}(t)$ with the impulse response $h(t)$ of the detector. 

In practice, detector bandwidth is set by the inverse of the period $t_\text{m}$ over which the signal is averaged:

\begin{eqnarray}
a^\text{m}_\text{out}(t) \approx \frac{1}{t_\text{m}}\int_{t-t_\text{m}/2}^{t+t_\text{m}/2} dt' e^{i \theta(t') - \kappa_\text{tot} t'/2}.
\end{eqnarray}
The resonant frequency $\omega_0$ doesn't appear in this expression because we assume that detection occurs after demodulation.

To evaluate the effect on the measured coherence time of the cavity, we focus on the measured average output field
\begin{eqnarray}
|\langle a^\text{m}_\text{out}\rangle| &=& \left|\langle\frac{1}{t_\text{m}}\int_{t-t_\text{m}/2}^{t+t_\text{m}/2}dt' e^{i\theta(t') - \kappa_\text{tot}/2}\rangle\right|. \nonumber
\end{eqnarray}
Expanding as in Ref.~\cite{martinis:2003},

\begin{eqnarray}
|\langle a^\text{m}_\text{out}\rangle|
&=&\left|\frac{1}{t_\text{m}}
\left[
\frac
{ \sinh(\kappa_\text{tot} t_\text{m}/2)}
{ (\kappa_\text{tot}/2)}
e^{-\kappa_\text{tot} t/2} 
- \right.\right.\\ \nonumber
&& \frac{1}{2} \omega_0^2 \int_0^\infty df S(f)  \times\\ \nonumber 
&& \ \left.\left.\int_{t-t_\text{m}/2}^{t+t_\text{m}/2}dt' e^{-\kappa_\text{tot} t'/2} W_0(f,t') + \mathcal{O}(<\theta(t')^3>)\right]\right|.
\label{T2}
\end{eqnarray}
Comparing this result to eq.~(\ref{eqn:T2fast}), we can see that the time-averaging has two effects.
The first effect is independent of the frequency fluctuations, as can be seen by setting $S(f) = 0$:
\begin{eqnarray}
|\langle a^\text{m}_\text{out}\rangle| 
=
\frac
{ \sinh(\kappa_\text{tot} t_\text{m}/2)}
{ (\kappa_\text{tot} t_\text{m}/2)}
e^{-\kappa_\text{tot} t/2}.
\label{T2nonoise}
\end{eqnarray}
In this limit, the time-dependence is an exponential decay characterized by a timescale $T_2=2/\kappa_\text{tot}$.  The averaging, however, can reduce the initial measured amplitude.  In the limit of fast measurements ($\kappa_\text{tot} t_\text{m} \ll 1$), unit amplitude is recovered.

The second effect of time averaging is a redefinition of the spectral weight function.  The new function is a weighted average of $W_0$ taken over the acquisition window.

By a similar argument, we can evaluate the effect of measurement bandwidth on the measured energy decay rate.
The average output power
evaluates to
\begin{eqnarray}
\langle |a^\text{m}_\textrm{out}|^2\rangle
&=& 
e^{-\kappa_\text{tot} t}\left(1-\omega_0^2\int_0^\infty df S(f) W_1(f) \right) \nonumber \\
&& +\mathcal{O}(\langle[\theta(t')-\theta(t'')]^3\rangle) +\mathcal{O}(\kappa_\text{tot} t_\text{m})^2. \nonumber
\label{amsf}
\end{eqnarray}
with
\begin{eqnarray}
W_1(\omega) &\equiv& \frac{1-\textrm{sinc}(\omega t_\text{m}/2)^2}{(\kappa_\text{tot}/2)^2+\omega^2}+\mathcal{O}(\kappa_\text{tot} t_\text{m})^3
\end{eqnarray}
and $\omega = 2\pi f$.

From Eq.~(\ref{amsf}), we see that the fluctuations change the contrast, but do not change the decay rate.  The amount by which they change the contrast is controlled by the spectrum of the fluctuations and the averaging time $t_\text{m}$ because the spectral weight $W_1(f)$ acts as a low-pass filter with corner frequency $1/t_\text{m}$. 

\subsection{Measurement of the cavity coupling rate}
\label{app:coupling}

After measuring the total loss rate $\kappa_{\text{tot}}$ of the system by fitting $\langle |a_{\text{out}}|^2\rangle$ to an exponential decay, measurement of the coupling loss rate $\kappa_{\text{ext}}$ enables the isolation of internal loss rate $\kappa_{\text{int}} = \kappa_{\text{tot}} - \kappa_{\text{ext}}$.
We now describe how the external coupling of the cavity  mode can be inferred from a ringdown experiment. Consider a cavity with classical mode amplitude $a$, frequency $\omega_\text{a}$, and total loss rate $\kappa_\text{tot}$ coupled to a single port with rate $\kappa_{\text{ext}}$.  The cavity may be driven through this port by an incident field $a_\text{in}$ with its carrier frequency $\omega_\text{d}$ detuned from the cavity by $\Delta\equiv\omega_\text{d}-\omega_\text{a}$.  In the rotating frame of the drive, the Langevin equation~\cite{collet:1983} is
\begin{eqnarray}
\dot{a} = (i\Delta -\kappa_\text{tot}/2)a + \sqrt{\kappa_{\text{ext}}} a_\text{in}.
\end{eqnarray}

\noindent The general solution for the cavity field is
\begin{eqnarray}
a(t) = -\sqrt{\kappa_{\text{ext}}} \int_0^t dt' a_\text{in}(t') e^{(i\Delta-\kappa_\text{tot}/2)(t-t')}.
\end{eqnarray}

Let the envelope of the incident field be that of a rectangular pulse,
 \[ a_\text{in} =\begin{cases} 
      0, & t~<0, \\ 
      a_0, & 0 < t< t_\text{p}, \\
      0, & t_\text{p} < t .
   \end{cases}
\]
The resulting cavity field is given by
 \[ a(t) =\begin{cases} 
      0, & t~<0, \\
      \tilde{a}(t), & 0~< t< t_\text{p}, \\
      \tilde{a}(t_\text{p})e^{(i\Delta-\kappa_\text{tot}/2)(t-t_\text{p})}, & t_\text{p} < t. 
   \end{cases}
\]
where 
\begin{eqnarray}
\tilde{a}(t) \equiv \frac{ a_0\sqrt{\kappa_{\text{ext}}}}{i\Delta -\kappa_\text{tot}/2}\left(1-e^{(i\Delta-\kappa_\text{tot}/2)t}\right).
\end{eqnarray}

We probe the cavity dynamics by measuring the emitted field $a_{\text{out}}$, which input-output theory relates to the incident field and the cavity mode~\cite{collet:1983}
\begin{eqnarray}
\label{eqn:input_output}
a_\text{out} = \sqrt{\kappa_{\text{ext}}} a + a_\text{in},
\end{eqnarray}
Substituting in the expressions for the input and cavity fields, 

\begin{eqnarray}
a_\text{out}(t) =\begin{cases} 
      0, & t~<0, \\
      \sqrt{\kappa_\text{ext}}\tilde{a}(t) + a_0, & 0< t< t_\text{p}, \\
      \sqrt{\kappa_{\text{ext}}}\tilde{a}(t_\text{p})e^{(i\Delta-\kappa_\text{tot}/2)(t-t_\text{p})}, & t_\text{p}< t .
   \end{cases}
\nonumber 
\\
\ 
\end{eqnarray}

The emitted fields slightly after the start of the pulse ($t=0^+$) and slightly after the end of the pulse ($t=t_\text{p}^+$) are:
\begin{eqnarray}
a_\text{out}(0^+) 
&=&
a_0, 
\nonumber 
\\
a_\text{out}(t_\text{p}^+)
&=& 
\sqrt{\kappa_{\text{ext}}} \tilde{a}(t_\text{p}).
\end{eqnarray}
These two points contain all the information required to extract the external coupling rate $\kappa_{\text{ext}}$.
The output field $a_{\text{out}}(t)$ at these points in time can be converted to phase-insensitive output powers $\langle |a_\text{out}|^{2}\rangle$. Assuming the cavity has been driven on-resonance ($\Delta = 0$),
\begin{eqnarray}
\label{eqn:P1andA1}
\langle |a_\text{out}(0^+)|^{2}\rangle 
&=& 
a^2_0,
\\
\langle |a_\text{out}(t_\text{p}^+)|^{2}\rangle 
&=& 
\frac{4 a^2_0 \kappa_{\text{ext}}^2}{\kappa_\text{tot}^2}
\left(
1 - 2e^{-\kappa_\text{tot} t_\text{p}/2} + e^{-\kappa_\text{tot} t_\text{p}}
\right).\nonumber
\end{eqnarray}

\noindent
Solving for the external coupling $\kappa_{\text{ext}}$ in terms of measured quantities and the experimental parameter pulse length $t_\text{p}$,

\begin{eqnarray}
\kappa_{\text{ext}} 
&=&
\frac
{\kappa_\text{tot}}
{2\left(1 - e^{-\kappa_\text{tot} t_\text{p}/2}\right)}
\sqrt{
\frac
{\langle |a_\text{out}(t_\text{p}^+)|^{2}\rangle}
{\langle |a_\text{out}(0^+)|^{2}\rangle}
}.
\end{eqnarray}

As the pulse is typically very short compared to $1/\kappa_\text{tot}$, we can instead expand Eq.~(\ref{eqn:P1andA1}) to second order in $\kappa_\text{tot} t_\text{p}/2$,

\begin{eqnarray}
\label{eqn:P1andA1_order2}
\langle |a_\text{out}(0^+)|^{2}\rangle 
&=& 
a^2_0,
\nonumber
\\
\langle |a_\text{out}(t_\text{p}^+)|^{2}\rangle 
&\approx& 
a^2_0 \kappa_{\text{ext}}^2 t_\text{p}^2,
\end{eqnarray}
and solve~(\ref{eqn:P1andA1_order2}) for $\kappa_{\text{ext}}$,
\begin{eqnarray}
\kappa_{\text{ext}} 
&\approx&
\frac
{1}
{t_\text{p}}
\sqrt{
\frac
{\langle |a_\text{out}(t_\text{p}^+)|^{2}\rangle}
{\langle |a_\text{out}(0^+)|^{2}\rangle}
}.
\end{eqnarray}

Although the output field at two precise moments in time is the only information needed for calculating $\kappa_{\text{ext}}$, a more precise determination of $\langle |a_\text{out}(t_\text{p}^+)|^{2}\rangle$ can be had by fitting the entire ringdown portion of the data ($t < t_\text{p}$) to an exponential decay and extracting the ringdown amplitude.

A similar strategy could be employed to improve the precision of $\langle |a_\text{out}(0^+)|^{2}\rangle$, but not as lucratively; because jitter in the system requires the pulse to be short (see App.~\ref{sec:nbar}), there is not much pulse data ($0<t<t_\text{p}$) for the fit. A workaround comes from creating a scenario where interference from cavity emission is negligible, thus resulting in $\langle |a_\text{out}(t)|^{2}\rangle$ having a constant amplitude $\langle |a_\text{out}(0^+)|^{2}\rangle$, even for an arbitrarily long pulse. Assuming the transmission background of the entire measurement system $|S_{21}|$ is smooth and varies only slowly with frequency, $\langle| a_\text{out}(0^+)|^{2}\rangle$ can be determined by measuring the transmitted power of a long pulse far above (perhaps hundreds of linewidths) and equally far below the cavity frequency $\omega_\text{a}$, then averaging the two for an approximation of $\langle |a_\text{out}(0^+)|^{2}\rangle$ at $\omega_\text{a}$. Pulses played far off-resonant from the cavity will not excite the cavity, so there will be no emission signal to interfere with the reflected pulse signal. 

If the cavity is frequency-tunable, as it is in the dipper by movement of the sample, then the off-resonance condition can be met by changing the frequency of the cavity instead of the frequency of the pulse. If the cavity is moved away from the frequency of interest, then a single pulse at that frequency can be used to measure $\langle |a_\text{out}(0^+)|^{2}\rangle$ at that frequency directly, rather than estimating it by linear interpolation between two measurements at other frequencies.  In practice this is the approach we take, although both methods give similar results.

Finally, the short pulses ($t_\text{p} \approx 5~\upmu \text{s}$) and weak coupling ($Q_\text{ext} \approx 2\times 10^8$; coupling rate $\kappa_\text{ext}/2\pi \approx 100~\text{Hz}$) used in this measurement mean that, for the same pulse amplitude, $\langle |a_\text{out}(0^+)|^{2}\rangle$ will be larger than $\langle |a_\text{out}(t_\text{p}^+)|^{2}\rangle$ by a few orders of magnitude. To avoid overdriving the receiver, the transmission measurements of $\langle |a_\text{out}(0^+)|^{2}\rangle$ can be performed with a pulse amplitude smaller by some reduction factor to make the $\langle |a_\text{out}(0^+)|^{2}\rangle$ measurements more comparable in size to the $\langle |a_\text{out}(t_\text{p}^+)|^{2}\rangle$ measurement. Before using this reduced-pulse measurement to calculate $Q_\text{ext}$, its result must be re-scaled such as to compensate for not only that reduction factor, but also for any signal compression on the way to the cavity.  Even if this compression does not affect this smaller pulse for measuring $\langle |a_\text{out}(0^+)|^{2}\rangle$, it may still affect the larger pulse used for measuring $\langle |a_\text{out}(t_\text{p}^+)|^{2}\rangle$.

\subsection{Calculating photon number}
\label{sec:nbar}

A cavity state with mode amplitude $a$ has an average of $n$ photons:
\begin{eqnarray}
\label{eqn:nbar_a}
n(t) = |a(t)|^2.
\end{eqnarray}
To calculate $n$, we can use Eq.~(\ref{eqn:input_output}) which relates the mode amplitude $a$ to more directly accessible quantities $a_\text{in}(t)$ or $a_\text{out}(t)$. To infer cavity photon number from the field emitted by the cavity, we can start by solving Eq.~(\ref{eqn:input_output}) for $a$ after the pulse has stopped ($a_\text{in}(t)=0$):
\begin{eqnarray}
a(t) = \frac{a_\text{out}(t)}{\sqrt{\kappa_\text{ext}}}.
\end{eqnarray}
Substituting into eq.~\ref{eqn:nbar_a} gives an expression for $n(t)$ in terms of the output field:
\begin{eqnarray}
\label{eqn:nbar_aout}
n(t) = \frac{|a_\text{out}(t)|^2}{\kappa_\text{ext}}.
\end{eqnarray}
This output-field equation for photon number can alternately be expressed in terms of an output power $P_\text{out}(t) = \hbar \omega_\text{a} |a_\text{out}(t)|^2$.  Solving for $n(t)$, we have
\begin{eqnarray}
n(t) 
=
\frac
{P_\text{out}(t) Q_\text{ext}}
{\hbar \omega_\text{a}^2}.
\end{eqnarray}

Using eq.~\ref{eqn:P1andA1}, the output-field equation~(\ref{eqn:nbar_aout}) can be converted into an input-field equation that predicts the number of photons injected by a rectangular excitation pulse of duration $t_\text{p}$ and amplitude $a_0$:
\begin{eqnarray}
\label{eqn:nbar_ain}
n(t_\text{p}) 
=
\frac{4 \kappa_\text{ext} a^2_0}{\kappa_\text{tot}^2}
\left(
1 - 2e^{-\kappa_\text{tot} t_\text{p}/2} + e^{-\kappa_\text{tot} t_\text{p}}
\right).
\end{eqnarray}
In the limit of a short pulse,
\begin{eqnarray}
n(t_\text{p}) 
=
\kappa_\text{ext} a^2_0 t_\text{p}^2 + \mathcal{O}\left(\kappa_\text{tot} t_p\right)^2.
\label{eqn:nbar_ain_short}
\end{eqnarray}
This input-field equation for photon number can alternately be expressed in terms of an input power $P_\text{in} = \hbar \omega_\text{d} |a_\text{in}|^2 = \hbar \omega_\text{d} a_0^2$, which allows Eq.~(\ref{eqn:nbar_ain_short}) to be expressed as
\begin{eqnarray}
n(t_\text{p}) 
=
\frac
{P_\text{in} t_\text{p}^2}
{\hbar Q_\text{ext}}.
\end{eqnarray}

So far, in this section we have assumed no jitter in the cavity frequency. In addition to simplifying Eq.~(\ref{eqn:nbar_ain_short}), a short drive pulse is important for proper control of photon number in the presence of jitter. In general, the photon number will be dependent upon the inner product of the cavity spectrum and the pulse spectrum. If the pulse is short enough that its spectrum is approximately uniform over the jitter width of the cavity, then that inner product is approximated well by the above expressions (such considerations are also essential for accurate measurement of $Q_\text{ext}^{-1}$ per App.~\ref{app:coupling}).

\section{Participation simulations}
\label{app:participations}
To calculate the participation ratios $p_j$,  we use finite-element analysis software (Ansys HFSS) to solve for the eigenmodes of Maxwell's equations in our cavity. 
The participation ratios are defined in terms of material parameters and the field solutions $\mathbf{E}$ and $\mathbf{H}$ for the fundamental cavity mode.
We consider four participations in our model: the bulk dielectric participation $p_\text{bulk}$, the substrate-air surface dielectric participation $p_\text{SA}$, the metal-air surface dielectric participation $p_\text{MA}$, and the conductor participation $p_\text{cond}$ (equal to the kinetic inductance ratio in bulk superconductors \cite{meservey:1969}).  We define these participations as
\begin{eqnarray}
p_\text{bulk} &\equiv& \frac{U_\text{bulk}}{U_\text{tot}}, \nonumber \\
p_\text{SA} &\equiv& \frac{U_\text{SA}}{U_\text{tot}}, \nonumber \\
p_\text{MA} &\equiv& \frac{U_\text{MA}}{U_\text{tot}}, \nonumber \\
p_\text{cond} &\equiv& \frac{U_\text{cond}}{U_\text{tot}},
\label{eqn:p_ratios}
\end{eqnarray}
where the energies $U_j$ in Eq.~(\ref{eqn:p_ratios}) are defined as~\cite{reagor:thesis,wang:2015}
\begin{eqnarray}
\label{eqn:p_energies}
\nonumber 
\\
U_\text{tot} 
&\equiv&
\frac{1}{2}
\text{Re}
\left\{
\int_{V_\text{tot}} 
\mathbf{E}\cdot \mathbf{D}^{\ast}\ dV
\right\},
\nonumber
\\
&=&
\frac{1}{2}
\text{Re}
\left\{
\int_{V_\text{tot}} 
\mathbf{H}\cdot \mathbf{B}^{\ast}\ dV
\right\}
,
\nonumber
\\
\newline 
\newline 
\nonumber
\\
U_\text{bulk} 
&\equiv&
\frac{1}{2}
\text{Re}
\left\{
\int_{V_\text{sub}} 
\mathbf{E}\cdot \mathbf{D}^{\ast}\ dV
\right\}
,
\nonumber 
\\
U_\text{SA} 
&\equiv& 
\frac{1}{2}t_\text{SA}
\text{Re}
\left\{
\int_{S_{\text{sub}}} 
\mathbf{E} \cdot \mathbf{D}^{\ast}\ dA
\right\}
,
\nonumber 
\\
U_\text{MA} 
&\equiv&
\frac{1}{2}t_\text{MA}
\text{Re}
\left\{
\frac{\epsilon_\text{0}}{\epsilon_\text{MA}}
\int_{S_\text{cav}} 
\mathbf{E} \cdot \mathbf{D}^{\ast}\ dA
\right\}
,
\nonumber 
\\
U_\text{cond} 
&\equiv&
\frac{1}{2} \lambda_\text{L} 
\text{Re}
\left\{
\int_{S_\text{cav}} 
\mathbf{H}\cdot \mathbf{B}^{\ast}\ dA
\right\}
.
\end{eqnarray}
Here, $V_\text{sub}$ and $S_\text{sub}$ refer to the volume and surface of the dielectric substrate under study, $S_\text{cav}$ is the surface of the aluminum cavity, and $V_\text{tot}$ is the volume of the entire system.

When solving for the eigenmodes of our system, we account for the birefringence of sapphire by assigning a tensorial relative permittivity $\overleftrightarrow{\epsilon_\text{bulk}}$ to the sample region. We assume a literature value for relative permittivity of 11.35 parrallel to the $c$-axis, and 9.27 perpendicular to the $c$-axis~\cite{krupka:1999}. 

For energies in material interface regions; $U_\text{SA}$, $U_\text{MA}$ and $U_\text{cond}$, volume integrals have been reduced to surface integrals by assuming field uniformity over the depth of the interface and integrating over the depth to yield a corresponding length prefactor.
This simplification allows these interface regions with distinct material properties to be represented by interfaces between regions defined in software, rather than by regions of their own. This amounts to a significant saving in processing time, as accurate field solutions in high aspect ratio volumes requires fine meshing, which can be computationally intensive. By this simplification, interface regions are treated as perturbations on the system, as the assumed material properties and thickness of these interfaces do not feed back and affect the field solutions acquired by the simulation.

The participations quoted in this work assume the surface dielectric thickness parameters $t_\text{SA}$ and $t_\text{MA}$ to be $3$ nm. We emphasize that this is a choice made for ease of comparison with surface loss tangents on silicon~\cite{wenner:2011}, and is not based on any direct knowledge of the surface thicknesses in our system.
The London penetration depth $\lambda_\text{L}$ is set to $50$~nm, which we confirm to be approximately accurate by fitting temperature sweeps of the cavity resonance frequency and quality factor to Mattis-Bardeen theory (see Table~\ref{tab:MB} in App.~\ref{app:temperature}).

For calculation of $U_\text{SA}$, it is crucial to perform the integral to one side of the substrate-air (SA) interface (rather than directly over it) to avoid ambiguity about the permittivity assigned by the software. 
The expression for $U_\text{SA}$ found in eq.~(\ref{eqn:p_energies}) is able to take on its compact form because of two choices made in our analysis. For one, we chose to perform the integral over the substrate-side of the SA-interface. Additionally, we assume the permittivity of the sample surface $\epsilon_\text{SA}$ to be the same as that of the sample bulk $\epsilon_{bulk}$, in which the integration was performed. The expression for $U_\text{SA}$ can be generalized to relax this equal permittivity assumption. This is done by decomposing the integrand into its components normal and tangent to the SA interface and invoking continuity of $\mathbf{E}$ and $\mathbf{D}$ where appropriate:

\begin{eqnarray}
\label{eqn:p_SA_general}
\nonumber
\\
\mathbf{E} \cdot \mathbf{D}^{\ast}
&\rightarrow&
{E}_{\parallel}
\left(
\frac
{\epsilon_{\parallel}^\text{SA}}
{\epsilon_{\parallel}^\text{bulk}}
{D}_{\parallel}
\right)^{\ast}
+
\left(
\frac
{\epsilon_{\perp}^\text{bulk}}
{\epsilon_{\perp}^\text{SA}}
{E}_{\perp}
\right)
{D}_{\perp}^{\ast}
.
\end{eqnarray}
After this generalizing substitution, $U_\text{SA}$ is still expressed in terms of the fields on the substrate-side of the SA interface. However, similar attention to boundary conditions permits $U_\text{SA}$ to be alternately expressed in terms of the fields in the vacuum immediately outside the sample, in case it is preferable to evaluate the surface integral over that side of the SA interface instead. 

Unlike the SA interface, the MA interface exists at the outer boundary of the simulation. This means when calculating $U_\text{MA}$, there is no ambiguity in the value of permittivity assigned by the software; nor is there any choice about the side of the interface over which to perform the integration\textemdash the integral must be evaluated over the vacuum side. Again invoking continuity of $\mathbf{E}$ and $\mathbf{D}$, fields in the MA interface region, along with the resulting $U_\text{MA}$, can be expressed in terms of the fields in the vacuum, where the field solutions are available. Unlike for the SA interface, however, this transformation is simplified by the conductor at the MA interface boundary, which requires all electric fields to be normal to the surface. With no tangential term in the integrand, the permittivity factor multiplying the normal term can be factored out of the integral. The result is that $U_\text{MA}$ uniformly scales inversely with $\epsilon_\text{MA}$, which we assume to be equal to $10$.

\section{Measurement sensitivity in a polynomial basis}
\label{app:poly_sens}
\begin{figure}[!htb]
\begin{center}
\includegraphics[width=1\linewidth]{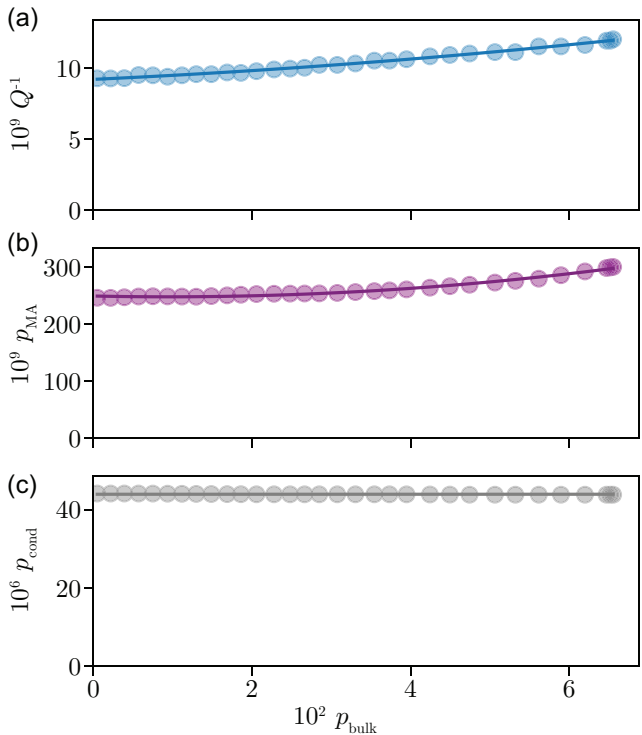}
\caption {\textbf{Polynomial representation}: (a)~the measured inverse internal quality factors, (b)~the simulated MA participation, and (c)~the simulated conductor participation of the HEMEX sample, each as a function of bulk participation.  All lines are 2nd-order polynomial fits per Eq.~(\ref{eqn:poly}).
}
\label{fig:polyfit}
\end{center}
\end{figure}

To gain intuition about what sets the sensitivities plotted in Fig.~\ref{fig:sensitivity}, we can use the mapping between $z$ and $p_{\text{bulk}}(z)$ (Fig.~\ref{fig:measurement_cartoon}b) to express $p_{\text{cond}}(z)$, $p_{\text{MA}}(z)$, and $Q^{-1}(z)$ as polynomials in $p_\text{bulk}$. In practice, 2nd-order polynomials closely approximate the system, as shown in Fig.~\ref{fig:polyfit}:
\begin{eqnarray}
Q^{-1} &=& y_0 + y_1 p_\text{bulk} + y_2 p_\text{bulk}^2, \nonumber \\ 
p_\text{cond} &=& x_0^\text{cond}, \nonumber \\
p_\text{bulk} &=& p_\text{bulk}, \nonumber \\
p_\text{MA} &=& x_0^\text{MA} + x_1^\text{MA} p_\text{bulk} +   x_2^\text{MA} p_\text{bulk}^2.
\label{eqn:poly}
\end{eqnarray}Recasting the problem in this fashion is convenient because conductor and substrate losses are orthogonal in the $p_\text{bulk}$-polynomial basis; the resulting matrix equation,  
\begin{eqnarray}
    \left(\begin{array}{c} 
    y_0 \\  
    y_1 \\
    y_2 \\
    \end{array}\right) = \left(\begin{array}{ccc} 
    x^{\text{cond}}_0 & 0 & x^{\text{MA}}_0\\  
    0 & 1 & x^{\text{MA}}_1\\  
    0 & 0 & x^{\text{MA}}_2\\
    \end{array}\right) \left(\begin{array}{c} 
    q_{\text{cond}}^{-1} \\  
    q_{\text{sub}}^{-1} \\
    q_{\text{MA}}^{-1} \\
    \end{array}\right),
    \label{polymat}
\end{eqnarray}
is nearly diagonal.

Conveniently, it yields a single equation for $q_\text{sub}^{-1}$:
\begin{eqnarray}
q_\text{sub}^{-1} = y_1 - x_1^\text{MA} q_\text{MA}^{-1}
\label{qsub}
\end{eqnarray}
Error propagation on Eq.~\ref{qsub} gives the desired expression for the sensitivity:

\begin{eqnarray}
\sigma_{q_\text{sub}^{-1}} &=& \sigma_{y_1} +  |x_1^\text{MA}|\sigma_{q_\text{MA}^{-1}} + \sigma_{x_1^\text{MA}}q_\text{MA}^{-1}.
\label{sigmaqsub}
\end{eqnarray}

The terms in Eq.~\ref{sigmaqsub} reveal what sets the measurement sensitivity, and how it can be improved.
To estimate the relative importance of the terms in Eq.~\ref{sigmaqsub}, we calculate their magnitudes assuming a hypothetical lossless substrate with the same permittivity as sapphire \cite{krupka:1999}.  Neglecting errors in alignment and simulation, the first term
$\sigma_{y_1}
\approx
\sqrt{\frac{1}{N}} 
\frac{\sigma_{Q^{-1}}}{\Delta p_\text{bulk}}
\approx 
1$ ppb.
The next two terms can be approximated using the coefficient $x_1^\text{MA}$ listed in Table~\ref{tab:poly}.  
\begin{table}[h!]
\centering
\begin{tabular}{|c || c | c c c|} 
 \hline
  $k$ & 
  $10^9~y_k$ & 
  $10^6~x_k^\text{cond}$ & 
  $x_k^\text{bulk}$ & 
  $10^9~x_k^\text{MA}$ \\ [0.5ex] 
 \hline
 0 & 9.18(0.05) & 43.92(0.01) & 0 & 249(1) \\ 
 1 & 26(3) & 0 & 1 & -300(60) \\
 2 & 240(40)  & 0 & 0& 15800(900) \\
 \hline
\end{tabular}
\caption{Polynomial coefficients $\mathbf{y}$ and $\mathbf{x}^j$ extracted from the fits to $Q^{-1}$ and $p_\text{j}$ in Fig.~\ref{fig:polyfit}.}
\label{tab:poly}
\end{table}
The two summands are then each approximately $2\times10^{-9}$.  If this analysis is repeated with polynomial order of 3 or 4, the $x_1^\text{MA}$ term grows to approximately 5 ppb.  Taken together, these considerations suggest that the sensitivity floor estimated from this polynomial analysis is several parts per billion, and is limited by our ability to separate MA losses from those caused by the substrate.  This baseline is consistent with the sensitivity shown in Fig.~\ref{fig:sensitivity}b.

This sensitivity of several parts per billion, when compared with the assumed measurement error of approximately 0.1 ppb indicates that input errors are amplified by a factor of roughly 100 during the regression.  This factor is much less than the condition number of the participation matrix, which is roughly $10^7$ for the samples considered in this work.  Such a discrepancy indicates the value of using the full covariance matrix.  

Note that the above argument neglects the 1\% fractional change in $p_\text{cond}$ and the 3\% fractional change in the cavity frequency that occur when the sample is inserted, which make conductive losses slightly dependent on insertion.  Accounting for these effects, such as by replacing  $p_\text{cond}$ with $\tilde{p}_\text{cond}$, results in a small but non-vanishing $x_1^\text{cond}$. Our analysis omits these effects as they change the cavity loss $Q^{-1}$ by less than $10^{-10}$.

\section{Sample preparation}
\label{app:provenance}
All samples are sourced from double-side-polished $c$-plane wafers (i.e., the extraordinary axis normal to the plane of the wafer). The EFG wafers are sourced from Kyocera, and the HEMEX wafers are sourced from GT Advanced Technologies (now called Crystal Systems). Samples are cut from the wafers, using a LatticeGear FlexScribe, to have dimensions of 6 mm x 45 mm. The samples are then cleaned with isopropanol and blown dry with nitrogen gas. Sample dimensions are verified by a wafer micrometer and Zeta-20 optical profilometer.

\section{Power sweeps}

\subsection{Defining the low-power regime}
\label{app:lp_regime}

We define the low-power regime as the range of cavity photon number $n$ which is sufficiently small such that total loss in the system $Q^{-1}$ is nearly desaturated and approximately the same as it would be at single-photon powers $Q^{-1}_{\text{lp}}$. This deviation of $Q^{-1}$ from single-photon behavior can be compared to the total saturable loss in the system to determine the saturation fraction $F$ of loss in the system:

\begin{eqnarray}
\label{eqn:saturationfraction}
F
\equiv
\frac{Q^{-1}_{\text{lp}} - Q^{-1}}{Q^{-1}_{\text{sat}}}
\end{eqnarray}

This value is identically zero when system behavior matches $Q^{-1}_{\text{lp}}$. Using Eq.~(\ref{eqn:TLS}), the saturation fraction can be expressed in terms of $n$, $n_\text{c}$ and $\alpha$.

\begin{eqnarray}
\label{eqn:satfracTLS}
F
=
1-\frac{1}{\sqrt{1+\left(\frac{n}{n_\text{c}}\right)^\alpha}}
\end{eqnarray}

Solving for $n$ indicates a maximum allowable photon number, given a maximum allowable saturation fraction and the parameters $n_\text{c}$ and $\alpha$ from measurement of the system.

\begin{eqnarray}
\label{eqn:lowpowerboundary}
\frac{n}{n_\text{c}}
<
\left(
\frac{1}{(1-F)^2}-1
\right)^{1/\alpha}
\end{eqnarray}

We define the boundary of the low power regime as the $n$ for which $F<0.1$. With $\alpha\approx 1/2$, this requires $n/n_\text{c} <5\times10^{-2}$. (Smaller values of $\alpha$ put less stringent bounds on $n$).  To be conservative, we choose $n=10^5$ such that $n/n_\text{c} < 10^{-3}$. Low-power measurements are made with a Josephson travelling wave parametric amplifier~\cite{macklin:2015}. 

\subsection{Power sweep bulk participations}
Table~\ref{tab:power_sweep_participations} shows the bulk participations for the inserted and withdrawn power sweeps on all samples.
\begin{table}[htb]
\begin{tabular}{ |c||c|c| } 
 \hline
 Material & withdrawn $p_\text{bulk}$ & inserted $p_\text{bulk}$ \\ \hline
 100\micron~EFG & $3.6\times 10^{-5}$ & $1.7\times 10^{-2}$ \\
 460\micron~EFG & $1.2\times 10^{-4}$ & $5.66\times 10^{-2}$ \\ 
 440\micron~HEMEX & $4.9\times 10^{-4}$ & $7.1\times 10^{-2}$ \\ \hline
\end{tabular}
\caption {\textbf{Bulk participations for inserted and withdrawn power sweeps}}
\label{tab:power_sweep_participations}
\end{table}

For power sweeps of the EFG samples, these $p_\text{bulk}$ are chosen to match that of the endpoints of the position sweeps in Fig.~\ref{fig:insertion_sweep}. This is done for ease of cross-comparison of power sweeps and position sweeps (as both have measurements with equal $p_\text{bulk}$ and $n = 10^5$). 

For HEMEX (measured by an earlier version of our protocol), we measured the inserted power sweeps with the sample in contact with the post, rather than some small distance ($z\approx 30$\micron) away from the post. This was done with the intent of maximizing the sensitivity to the sample loss, but had the drawback of removing direct cross-comparability between the power sweep and position sweep of HEMEX. Based on the trend in $Q^{-1}$ measured as a function of $p_\text{bulk}$ (Fig.~\ref{fig:insertion_sweep}), the large participation of the inserted power sweep is insufficient to explain the large $Q^{-1}$ measured at the lowest excitation powers (Fig.~\ref{fig:pow_sweep}).  We posit that this discrepancy was caused by measuring the inserted HEMEX power sweep with the sample in contact with the top of the post, which could make the cavity mode more sensitive to surface losses associated with surface roughness or material non-homogeneity, neither of which are captured by our simulations. 

The different condition under which HEMEX was measured does not invalidate the low-power verification done by the inserted power sweep, as further insertion of the sample tends to reduce the critical photon number $n_\text{c}$. This can be seen by comparing Table~\ref{tab:pow_sweep} to Table~\ref{tab:pow_sweep_w}. The result of this effect is to give a stricter criterion for the boundary of the low-power regime, making the verification process conservative.

\begin{table}[htb]
\begin{tabular}{ |c||c|c|c| } 
 \hline
 Sample  & $n_\text{c}$ (billions)& $\alpha$ \\ \hline
 $100$\micron~EFG   & $7(3)$ &  $0.5(0.1)$\\ 
  $460$\micron~EFG & $8(1)$ & $0.4(0.16)$\\
 $440$\micron~HEMEX & $14(4)$ & $0.32(0.03)$\\ \hline
\end{tabular}
\caption {Fit results for the withdrawn (orange) power sweeps in Fig.~(\ref{fig:pow_sweep}) using the TLS model in Eq.~(\ref{eqn:TLS}).} 
\label{tab:pow_sweep_w}
\end{table}

\subsection{Constraining the position sweep fit}
\label{app:ps_constraints}
When the sample is withdrawn, the total internal loss of the system $Q^{-1}$ consists predominantly of two loss channels: conductivity loss $Q^{-1}_\text{cond}$ and metal-air interface dielectric loss $Q^{-1}_\text{MA}$.  Power sweeps of the cavity provide some information about these individual underlying loss mechanisms. This information can be translated into bounds on the low-power material loss factors $q^{-1}_{j}$ associated with the cavity, and those bounds can be applied to the analysis of the position sweep described in Sec.~\ref{sec:Measurement technique}.

As with any resonator measurement, $Q^{-1}$ can act as an upper bound on the loss of each underlying loss channel. For instance, the low-power value of $Q^{-1}_\text{MA}$ can be bounded from above by the low-power total $Q^{-1}$, and that upper-bound can be converted into a bound on the low-power value of $q^{-1}_\text{MA}$. 
Additional information about $Q^{-1}_\text{MA}$ comes from assuming that conductivity loss $Q^{-1}_\text{cond}$ is power-independent up to $n=10^{14}$ (an assumption consistent with observations up to $n=10^{10}$ in prior studies of aluminum cylindrical cavities \cite{reagor:2013}). 
If $Q^{-1}_\text{cond}$ is power-independent, then any power-dependence observed in $Q^{-1}$ must be attributed to $Q^{-1}_\text{MA}$, and this attribution can be converted into a lower-bound on $q^{-1}_\text{MA}$ (see eq.~\ref{eq:psweep_bounds}).

An upper-bound on $Q^{-1}_\text{cond}$ can be inferred in the same way, but assuming that $Q^{-1}_\text{cond}$ is power-independent implies a slightly stricter bound. If the low-power and high-power values of $Q^{-1}_\text{cond}$ are assumed to be the same, then  the high-power $Q^{-1}$ (or any power $Q^{-1}$) can serve as an upper-bound on the low-power $Q^{-1}_\text{cond}$, and converted into an upper-bound on $q^{-1}_\text{cond}$ (see eq.~\ref{eq:psweep_bounds}).

The precise bounds obtained for each sample are listed in Table~\ref{tab:cav_loss_bounds}.
These bounds are obtained separately for each cooldown, to control for potential temporal variations in the bare quality factor of the cavity.  Over the course of the measurements presented in this study, however, no such variations were observed. Consequently, the bounds in Table~\ref{tab:cav_loss_bounds} are relatively similar for all three measurements.

\begin{table}[htb]
\begin{tabular}{ |c||c| } 
 \hline
 Sample material & Cavity loss factor bounds\\ 
 \hline
 
100\micron~EFG 
& $8.62\times10^{-3} \leq q_\text{MA}^{-1} \leq 38.2\times10^{-3}$ \\ 
& $\hspace{1.45cm} 0\leq q_\text{cond}^{-1}\leq 1.66\times10^{-4}$ \\ 

460\micron~EFG 
& $9.18\times10^{-3}\leq q_\text{MA}^{-1} \leq 37.6\times10^{-3} $ \\ 
& $\hspace{1.45cm} 0\leq q_\text{cond}^{-1}\leq 1.60\times10^{-4}$ \\ 

440\micron~HEMEX 
& $9.33\times10^{-3}\leq q_\text{MA}^{-1} \leq 38.3\times10^{-3} $ \\ 
& $\hspace{1.45cm} 0\leq q_\text{cond}^{-1}\leq 1.62\times10^{-4}$  \\ 

\hline
\end{tabular}
\caption {\textbf{Constraints on cavity loss factors}}
\label{tab:cav_loss_bounds}
\end{table}

When performing the insertion analysis in Fig.~\ref{fig:insertion_sweep}, the three measurements are fit jointly, which has the advantage that it removes four free parameters from the fit, reducing the risk of over-fitting. 
The ``background'' loss rates extracted in this way are shown in Table~\ref{tab:fit_results}, and are consistent with other measurements of etched 4N aluminum cavities~\cite{reagor:2013,axline:thesis}.  (The quoted value for conductor loss is given at the withdrawn frequency of 4.55 GHz.) We could repeat the insertion analysis with the cavity loss factors constrained using the bounds from Table~\ref{tab:cav_loss_bounds}, but this has no affect on the fit, as the unconstrained fit already yields loss rates that fall within the bounds in  Table~\ref{tab:cav_loss_bounds}. This demonstrates consistency between the power analysis from Sec.~\ref{sec:results}A and the insertion analysis from Sec.~\ref{sec:results}B.

\begin{table}[htb]
\begin{tabular}{ |c||c|c| } 
 \hline
 Loss factor & Extracted value\\ \hline
 $q^{-1}_\text{MA}$ & $33(9)\times10^{-3}$ \\ 
 $q^{-1}_\text{cond}$ & $2(5)\times10^{-5}$ \\ 
  $R_\text{s}$ & $4(9)\times10^{-8} ~\Omega$ \\ 
 $q^{-1}_\text{sub}$ (thin EFG) & $170(13)\times10^{-9}$\\ 
  $q^{-1}_\text{sub}$ (thick EFG) & $88(6)\times10^{-9}$\\ 
   $q^{-1}_\text{sub}$ (HEMEX) & $19(6)\times10^{-9}$\\
   $q^{-1}_\text{bulk}$ (EFG) & $63(8)\times10^{-9}$\\
   $q^{-1}_\text{SA}$ (EFG) & $15(3)\times10^{-4}$\\\hline
\end{tabular}
\caption {Summary of fit results.} 
\label{tab:fit_results}
\end{table}

\section{Temperature dependence}
\label{app:temperature}

Conductor loss from the aluminum walls of the cavity is temperature dependent. Near the vicinity of the superconducting transition temperature $T_\text{c}$, it is much greater than the metal-air (MA) surface dielectric loss from the cavity oxide layer.  We utilize this fact to experimentally confirm that insertion of the dielectric does not change the conductor participation of the cavity mode.

\begin{figure*}[!tb]
\begin{center}
\includegraphics[width=1\linewidth]{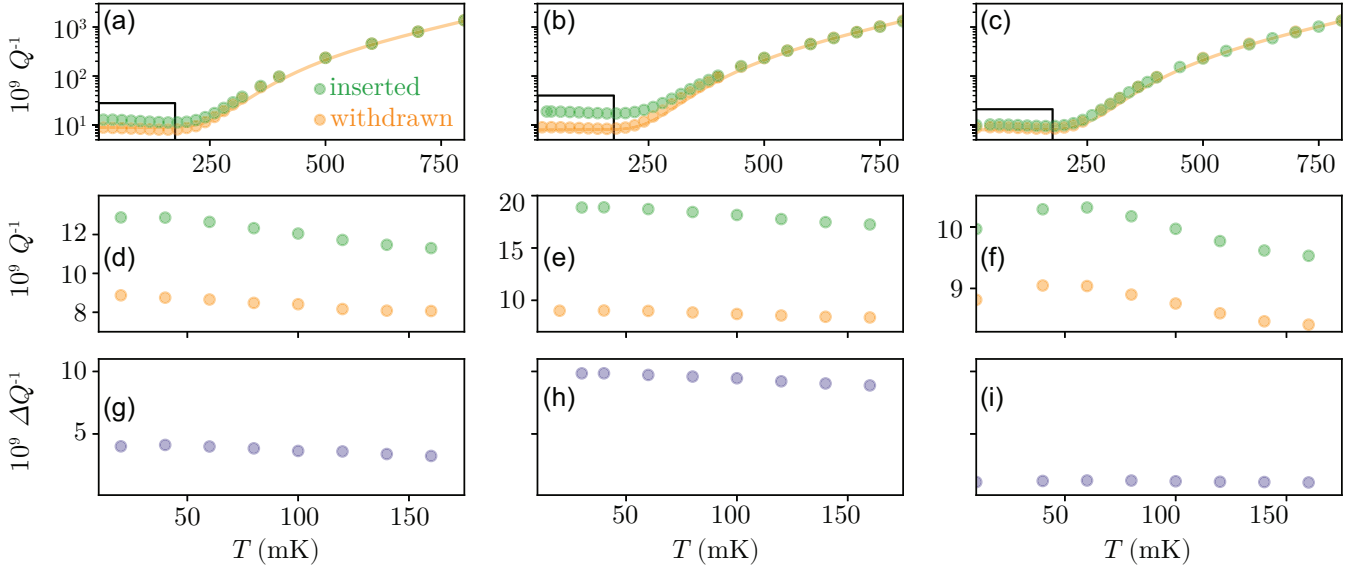}
\caption {\textbf{Temperature sweeps}. Measured internal loss $Q^{-1}$ with the sample withdrawn (orange) and inserted (green) as a function of the mixing-chamber plate temperature, for (a)~$100$\micron~thick EFG sapphire, (b)~$460$\micron~thick EFG sapphire, and (c)~$440$\micron~thick HEMEX sapphire. Orange lines are fits to Mattis-Bardeen theory~\cite{mattis:1958}, as described in the supplement of Ref.~\cite{lei:2020}.
}
\label{fig:temp_sweep}
\end{center}
\end{figure*}

Fig.~\ref{fig:temp_sweep} shows measurements of cavity loss rate $Q^{-1}$ as a function of the temperature $T$ measured at the mixing chamber plate of the dilution refrigerator.  Measurements of $Q^{-1}$ are performed with the sample fully withdrawn (orange traces) and fully inserted (green traces).  

Two distinct regimes are visible in the measurements. For temperatures at or near $T_\text{c}$, conductor losses dominate, and system behavior can be described by Mattis-Bardeen theory~\cite{mattis:1958}, which we fit to the temperature dependence of the withdrawn temperature sweeps (orange traces). The results are shown in Table~\ref{tab:MB}.

\begin{table}[htb]
\begin{tabular}{ |c|c||c|c|c| } 
 \hline
 Material & Thickness & $T_\text{c}$ (K) & $\lambda_\text{L}$ (nm) \\ \hline
 EFG sapphire & $100$\micron & $1.13$ &  $135$\\ 
 EFG sapphire & $460$\micron & $1.19$ & $90$\\
 HEMEX sapphire & $440$\micron & $1.13$ & $90$\\ \hline
\end{tabular}
\caption {Cavity material properties according to fit results for withdrawn (orange) temperature sweeps in Fig.~(\ref{fig:temp_sweep}) using Mattis-Bardeen theory~\cite{mattis:1958}.} 
\label{tab:MB}
\end{table}

Near $T_\text{c}$, conductivity loss grows to be orders of magnitude larger than the other loss channels, making total loss approximately proportional to participation in the conductor: $Q^{-1}\approx p_\text{cond} q_\text{cond}^{-1}$. A pair of high-temperature measurements with the sample withdrawn and inserted may therefore serve as a probe of the fractional change in $p_\text{cond}$ at the two measured positions. 

\begin{eqnarray}
\label{eqn:pcondcheck}
\frac
{1/Q^{\text{I}}-1/Q^{\text{W}}}
{1/Q^{\text{W}}} 
\approx 
\frac
{p_{\text{cond}}^{\text{I}}-p_{\text{cond}}^{\text{W}}}
{p_{\text{cond}}^{\text{W}}}.
\end{eqnarray}
(Here the superscripts indicate whether a measurement is made with the sample withdrawn or inserted).  With this argument, high-temperature measurements with the sample withdrawn and inserted bound the fractional change in conductor participation to less than 1\%, in agreement with the simulations shown in App.~\ref{app:poly_sens}.

For temperatures $T\ll T_\text{c}$, conductor losses are only weakly dependent upon temperature, so we attribute the observed temperature dependence to the dielectrics in the system.  Fig.~\ref{fig:temp_sweep}d-f shows this regime in greater detail.  Weak temperature dependence is visible in both the inserted and withdrawn sweeps, indicating some saturability of the dielectric losses.  However, subtracting the withdrawn and inserted measurements, as shown in Fig.~\ref{fig:temp_sweep}g-i, reveals that inserting the sample does little to change this temperature dependence, indicating that the temperature dependence arises primarily from the metal-air interface of the cavity, rather than from the sample under study.  In the future, modifications to the experimental design could improve the contrast of such temperature-dependence comparisons by better thermalizing the dielectric substrate to the mixing chamber, or by constructing the cavity from a superconductor with higher transition temperatures, such as niobium.

\section{Repeatability and cross-wafer variation}
\label{app:repeatability}
\begin{figure}[!htb]
\begin{center}
\includegraphics[width=1\linewidth]{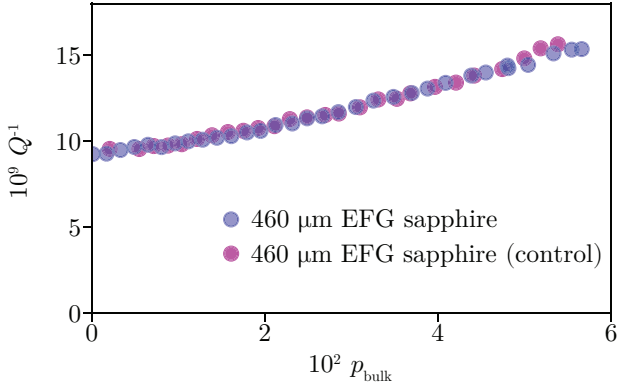}
\caption {\textbf{Position sweep on nominally identical samples}. The position sweep in Fig.~\ref{fig:insertion_sweep} was repeated with a second piece of $460$\micron~thick EFG sapphire denoted as a control. The control was prepared in nominally the same way, but cut from a different wafer of the same material.
}
\label{fig:control}
\end{center}
\end{figure}

As a control test to check for measurement reproducibility and wafer-to-wafer variation in nominally identical samples, we repeated the position sweep of $460$\micron~thick EFG sapphire on a second control sample cut from a different wafer. The results are shown in Fig.~\ref{fig:control}. 
The two measurements are quite similar, suggesting that the dipper is a measurement technique with high reproducibility and that the dielectric properties of these two samples are the same, up to the resolution of the measurement.

\section{Bounding magnetic bulk loss}
\label{app:mag_loss}
Implicit in our analysis is the assumption that the origin of loss from the sample is purely dielectric and none of the loss is magnetic. In this appendix we relax that assumption, and modify our model to consider the possibility that some of the loss from the substrate could be magnetic in origin. We then compare our measurement results of EFG sapphire with quality factors of stripline resonators fabricated on the same material. From this comparison, we can make bounding statements about the composition of bulk loss in our sample, and about the prevalence of bulk magnetic loss in superconducting qubits on EFG sapphire.

Just as substrate loss can be decomposed into bulk and surface contributions (See Eq.~\ref{eqn:bulk/SA separation} and Fig.~\ref{fig:insertion_sweep}b), we can further decompose loss from the bulk of our sample $Q_\text{bulk}^{-1}$ into contributions from bulk dielectric loss $Q_\text{bulk(E)}^{-1}$ and bulk magnetic loss $Q_\text{bulk(H)}^{-1}$. These losses can be treated with the same participation formalism and attributed to an associated dielectric loss tangent $q_\text{bulk(E)}^{-1}$  and a magnetic loss tangent $q_\text{bulk(H)}^{-1}$, respectively:
\begin{eqnarray}
\label{eqn:bulk_diel_mag}
Q_\text{bulk}^{-1} 
=
p_\text{bulk(E)}  q_\text{bulk(E)}^{-1}
+
p_\text{bulk(H)}  q_\text{bulk(H)}^{-1}
.
\end{eqnarray}

\begin{figure}[!htb]
\begin{center}
\includegraphics[width=1\linewidth]{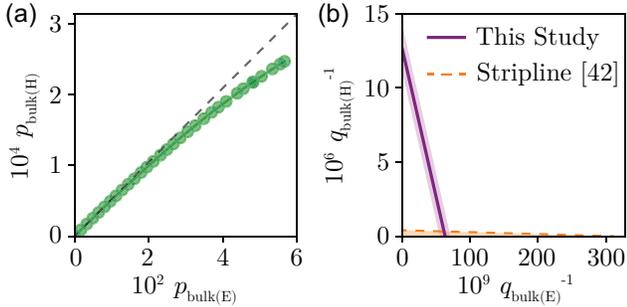}
\caption {\textbf{Bounding magnetic bulk loss}. (a) Bulk magnetic participation $p_\text{bulk(H)}$ as a function of bulk dielectric participation $p_\text{bulk(E)}$. (b) Contours representing viable decompositions of bulk substrate loss into dielectric and magnetic components in our experiment (Eq.~\ref{eqn:bulk_diel_mag}) and in a stripline resonator~\cite{axline:2016} (Eq.~\ref{eqn:bulk_diel_mag_stripline}). After our model is generalized to allow losses from the substrate bulk to be a combination of dielectric and magnetic loss (purple contour), the relative contributions of these losses can be constrained by the performance of a stripline fabricated on the same substrate (orange bounding contour). To be consistent with stripline $Q$, it must be that at least $97\%$ of the sample loss observed in our measurement must be dielectric in origin. These constraints also suggest that magnetic loss is at most a $10\%$ contributor to the decoherence of transmons on EFG sapphire.}
\label{fig:mag_loss}
\end{center}
\end{figure}

Because of the quasi-linear relationship between bulk dielectric participation $p_\text{bulk(E)}$ and bulk magnetic participation $p_\text{bulk(H)}$ in our geometry (see Fig.~\ref{fig:mag_loss}a), our measurement has limited ability to distinguish between these two loss channels but it does constrain them to a contour in the space of all ordered pairs $(q_\text{bulk(E)}^{-1},q_\text{bulk(H)}^{-1})$. 
In close analogy to the method of bulk-surface loss separation performed in Fig.~\ref{fig:insertion_sweep}b, an additional measurement with a different ratio of participations $p_\text{bulk(E)}/p_\text{bulk(H)}$ would suggest another constraint line with a different slope, and further constrain the accessible $(q_\text{bulk(E)}^{-1},q_\text{bulk(H)}^{-1})$ pairs. Because our measurements probe loss in our sample with the fringe field of a charge antinode, our measurement is preferentially sensitive to dielectric loss, and achieves $p_\text{bulk(E)}/p_\text{bulk(H)} \approx 200$. To acquire another constraint line, we can consider the performance of a stripline fabricated on EFG sapphire from another study~\cite{axline:2016}. A stripline of this design can achieve $Q \approx 8\times10^6$  and has $p_\text{bulk(E)}/p_\text{bulk(H)} = 0.40/0.31 \approx 1.3$; producing a contour with a much more gradual slope which we pair with our steeper measurement contour (See Fig.~\ref{fig:mag_loss}b). Because the $Q$ of the stripline is the result of a multitude of possible loss channels, the associated contour represents an inequality: 
\begin{eqnarray}
\label{eqn:bulk_diel_mag_stripline}
Q^{-1} 
\geq
p_\text{bulk(E)}  q_\text{bulk(E)}^{-1}
+
p_\text{bulk(H)}  q_\text{bulk(H)}^{-1}
.
\end{eqnarray}
This means that the intersection of the two contours cannot be used to extract the true value of $(q_\text{bulk(E)}^{-1}, q^{-1}_\text{bulk(H)})$  associated with our sample, however  upper and lower bounds can be inferred on both these loss tangents using the region where our measurement contour overlaps the region bounded from above by the stripline contour.

The contours bound the bulk dielectric loss tangent to $6.1 \times 10^{-8} \leq q_\text{bulk(E)}^{-1} \leq 6.3 \times 10^{-8}$, meaning that even considering the possibility that there may be magnetic losses associated with the sample bulk, this can account for no more than $3\%$ of the sample loss detected in our measurement, and at least $97\%$ of the bulk loss we detect must be dielectric in origin.  This validates the analysis in the main text, which neglects magnetic losses.
Although prior studies of HEMEX sapphire dielectric resonators have observed tunable loss due to coupling with electron spins of magnetic impurities brought into resonance by tesla-scale magnetic fields~\cite{benmessai:2013,farr:2013,goryachev:2014}, our result shows that dielectric loss is the dominant source of dissipation in EFG sapphire at low magnetic fields.

The contours also bound the bulk magnetic loss tangent to $q_\text{bulk(H)}^{-1} \leq 3.3 \times 10^{-7}$. This bound is much looser than the bounds for $q_\text{bulk(E)}^{-1}$, largely due to the relative insensitivity of our measurement to magnetic losses. Fortunately, a typical transmon qubit, because of its high kinetic inductance, has $p_\text{bulk(H)} \approx 2.5\%$ (relatively small compared to the 31\% of a typical stripline), so this bound on $q_\text{bulk(H)}^{-1}$  can still be used to infer a useful transmon coherence limit of $Q_\text{bulk(H)} >1.2 \times 10^8$. This high coherence limit suggests that even in today's longest-lived transmons on sapphire, bulk magnetic loss accounts for less than $10\%$ of total loss, making bulk magnetic loss at most a minor contributor to their decoherence.

\section{Comparison to other studies}
\label{app:lit_compare}
\begin{figure}[!hb]
\begin{center}
\includegraphics[width=1\linewidth]{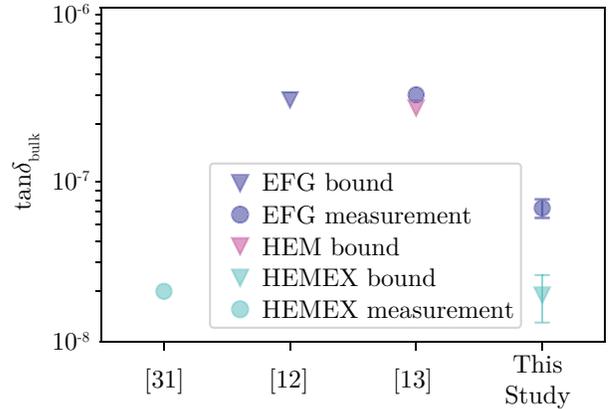}
\caption {\textbf{Comparison of low-power and low-temperature loss-tangent measurements of sapphire}.}
\label{fig:lit_compare}
\end{center}
\end{figure}

Fig.~\ref{fig:lit_compare} compares the loss tangents measured in this work to those reported in other studies on the microwave loss of sapphire at low temperatures and low powers.

\section{Microwave measurement circuit}
\label{app:circuit}
\begin{figure*}[!]
\begin{center}
\includegraphics[width=1\linewidth]
{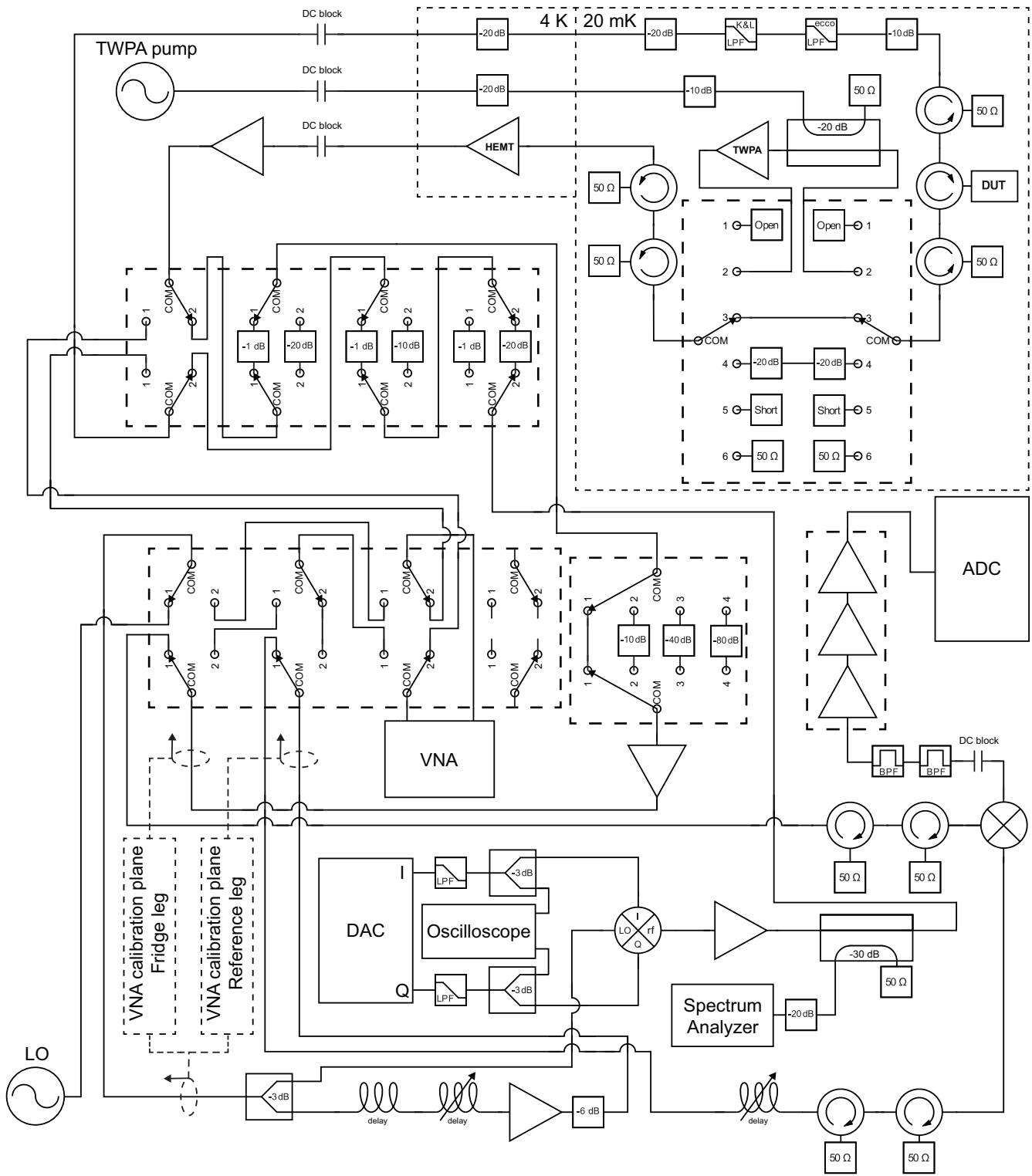}
\caption {\textbf{Microwave measurement circuit}. Microwave cavity (DUT) is cooled to $20$ mK and measured with heterodyne detection, employing switched amplifier and attenuator banks for improved dynamic range.
}
\label{fig:circuit}
\end{center}
\end{figure*}

Fig.~\ref{fig:circuit} shows the microwave wiring diagram of the measurement system.


\bibliographystyle{unsrt}
\bibliography{dipper_bib}

\begin{thebibliography}{10}
\bibitem{tSA}
We assume that surface dielectric interfaces have a thickness of 3 nm (see
  App.~\ref{app:participations}).
\end{thebibliography}

\end{document}